\documentclass[11pt]{article}

\pdfoutput=1 

\usepackage[dvipsnames]{xcolor}
\usepackage{amssymb,amsmath,amscd}
\usepackage{epsfig}
\usepackage{epstopdf}
\usepackage{latexsym}
\usepackage{graphicx}
\usepackage{booktabs}
\usepackage{bbm}
\usepackage[margin=15pt,small]{caption}
\usepackage{subcaption}
\usepackage{enumitem}
\usepackage{float}
\usepackage[toc]{appendix}
\usepackage[numbers,compress]{natbib}
\usepackage{tikz}
\usepackage[labelfont=bf]{caption}
\usepackage[export]{adjustbox}
\usepackage{nicematrix}
\usepackage{braket,cellspace}
\usepackage{mathrsfs}
\usepackage{braket}
\usepackage{chngpage}
\usetikzlibrary{positioning} 

\definecolor{darkspringgreen}{rgb}{0.09, 0.45, 0.27} 
\usepackage[
      colorlinks=true,
      linkcolor=darkspringgreen,  
      urlcolor=darkspringgreen,    
      filecolor=darkspringgreen,     
      citecolor=darkspringgreen,
      linktocpage=true,
      pdfstartview=FitV,
      bookmarksopen=true     
      ]{hyperref} 

\numberwithin{equation}{section} 
\allowdisplaybreaks 
\setlist{nolistsep} 

\let\oldbibliography\thebibliography
\renewcommand{\thebibliography}[1]{\oldbibliography{#1}
\setlength{\itemsep}{4pt}} 

\begin{document}  

\begin{titlepage}
\begin{center} 

\vspace*{20mm}

{\LARGE \bf 
A Solvable Model of Flat Space Holography
}

\vspace*{20mm}
 
{\bf Felipe Rosso \\ }
\vskip 2mm
Department of Physics and Astronomy 
\\
University of British Columbia \\
Vancouver, BC V6T 1Z1, Canada  

\bigskip
\tt{feliperosso6@gmail.com}  \\

\end{center}

\bigskip

\begin{abstract}
\noindent We propose an explicit realization of flat space holography in two dimensions where both sides of the duality are independently defined and the boundary theory is completely solvable. In the bulk, we define a novel $\mathcal{N}=1$ flat space supergravity theory and exactly compute the full topological expansion of its Euclidean partition function with an arbitrary number of boundaries. On the boundary, we consider a double scaled Hermitian random matrix model with Gaussian potential and use the loop equations to show it independently reproduces the bulk partition function to all orders in the topological expansion. The non-perturbative completion of the supergravity theory provided by the solvable Gaussian matrix model allows for the  exact, and in many cases analytic, computation of observables in flat space quantum gravity.
\end{abstract}

\vfill
\end{titlepage}

\setcounter{tocdepth}{2}
\tableofcontents

\section{Introduction}

The aim of this work is to make progress towards a non-perturbative definition of quantum gravity with vanishing cosmological constant. This is the ultimate goal of the celestial holography program \cite{Strominger:2017zoo,Raclariu:2021zjz,Pasterski:2021rjz}, whose focus is on the S-matrix of four-dimensional asymptotically flat gravity. The proposal is that the gravitational theory admits a holographic description in terms of a putative two-dimensional celestial conformal field theory, from which S-matrix elements can be extracted. Even though a great deal of progress has been made in this direction, an explicit realization of the holographic duality, where both sides are independently defined, is currently lacking.

Here, we take a different approach to the problem and instead focus on the much simpler case of two-dimensional gravity theories, hoping such toy models provide valuable general lessons about the nature of flat space quantum gravity (as has certainly been the case for AdS \cite{Saad:2019lba,Almheiri:2019psf,Penington:2019kki,Almheiri:2019qdq}). One of the simplest flat space theories is CJ gravity, introduced by Cangemi and Jackiw in the nineties \cite{Cangemi:1992bj}. Building on \cite{Afshar:2019tvp,Afshar:2019axx,Godet:2020xpk,Godet:2021cdl}, an exact holographic dual for this theory has been recently proposed in \cite{Kar:2022vqy,Kar:2022sdc}. Similarly as in other AdS$_2$ cases \cite{Saad:2019lba}, the holographic system is not given by a single celestial theory but instead an ensemble of random matrices, i.e. a celestial matrix model. 

In this article we define the natural $\mathcal{N}=1$ supersymmetric extension of CJ gravity and provide rigorous evidence that points to a double scaled Hermitian random matrix model as its holographic dual. Our construction is particularly appealing given that the matrix model is as simple as it can be: its probability density is characterized by a Gaussian potential. This gives a concrete realization of flat space holography in two-dimensions where the dual system is not only under complete control but where full non-perturbative effects can be studied, in many cases, analytically. In the remainder of the introduction we summarize and provide further technical details on the construction presented in the main text.

\paragraph{Summary of results:} We begin in Section \ref{sec:2}, which contains all the analysis regarding the gravitational side of the duality. Cangemi and Jackiw originally defined the CJ gravity action (\ref{eq:CJAction}) from a BF gauge theory built from the Maxwell algebra (\ref{eq:MaxwellAlgebra}), a central extension of the Poincaré algebra. In Subsection~\ref{sec:2.1} we follow their approach and construct the $\mathcal{N}=1$ CJ supergravity action (\ref{eq:SCJAction}) from a BF theory with an appropriate supersymmetric extension of the Maxwell algebra (\ref{eq:MaxwellSuperalgebra}). The bosonic field content of the theory is the same as in CJ gravity: the metric $g_{\mu \nu}$ and dilaton $\Phi$ are accompanied by the scalar $\Psi$ and the abelian topological gauge field $A_\mu$. The novelty is in the fermionic sector, given by a gravitino $\psi^\alpha_\mu$ and a dilatino $\lambda^\alpha$. Apart from its equivalence to a BF theory, the other crucial element of CJ supergravity which makes it a very tractable model, is the linear dependence of the dilaton in the action, given by
\begin{equation}\label{eq:IntroLinearDilaton}
I_{\rm SCJ} \supset \frac{1}{2}\int_\mathcal{M} d^2x\sqrt{g}\,\Phi R\ .
\end{equation}
Varying with respect to $\Phi$ forces all solutions of the theory to be locally flat.

In Subsection~\ref{sec:2.2} we study the physical degrees of freedom of the theory which, due to its topological nature, are completely localized at the boundary. After prescribing a natural set of asymptotic boundary conditions, we determine the infinite dimensional superalgebra (\ref{eq:AsymptoticAlgebra1}) satisfied by large gauge transformations and show that (after a redefinition of its generators) it is equivalent to the three dimensional BMS superalgebra (\ref{eq:sBMS3}) of \cite{Barnich:2014cwa}. We continue in Subsection \ref{sec:2.3} where we find the dynamics of the boundary modes are controlled by a simple $\mathcal{N}=1$ supersymmetric quantum mechanics with the following action\footnote{The inverse length scale $\gamma$ is introduced via the boundary conditions, see footnote \ref{foot:gamma}.}
\begin{equation}\label{eq:IntroBounAct}
I_\partial[F,h,\vartheta]=\gamma\int_0^\beta 
\frac{d\tau}{F'(\tau)}\left[h'(\tau)F''(\tau)+2\vartheta'(\tau)\vartheta''(\tau)\right]\ ,
\end{equation}
where $\big(F(\tau),h(\tau)\big)$ are $\beta$-periodic functions and $\vartheta(\tau)$ an anti-periodic Grassmann function. The Euclidean coordinate $\tau\sim \tau+\beta$ parametrizes the boundary of a flat manifold with the topology of the disk, obtained by analytically continuing the retarded Bondi time $u\rightarrow i\tau$. The bosonic sector of this action agrees with the one obtained for ordinary CJ gravity in \cite{Afshar:2019axx}.

Having defined and understood the degrees of freedom of the theory, in Subsection \ref{sec:2.4} we compute the observable we are mostly interested in: the Euclidean partition function $Z(\beta_1,\dots,\beta_n)$ with an arbitrary number of boundaries with periodicity $\beta_i$. In two dimensions, the partition function admits the following topological expansion
\begin{equation}\label{eq:IntroTopological}
Z(\beta_1,\dots,\beta_n)\simeq \sum_{g=0}^{\infty}(e^{-S_0})^{2(g-1)+n}Z_g(\beta_1,\dots,\beta_n)\ ,
\end{equation}
with $S_0\in \mathbb{R}_+$, the parameter multiplying the Euler characteristic term in the action, and $Z_g$ the partition function which only includes contributions from surfaces of fixed genus $g$. The symbol~$\simeq$ reminds us that such expansion is not the full result as it is missing, by construction, doubly non-perturbative corrections $\mathcal{O}(e^{-e^{S_0}})$. 

There are two features that allow us to compute this expansion exactly. First, since $\Phi$ appears linearly in the action (\ref{eq:IntroLinearDilaton}) we can solve the dilaton path integral along a purely imaginary contour and obtain a Dirac delta $\delta(R)$, meaning only locally flat surfaces with asymptotic boundaries actually contribute to the partition function. The classification of orientable locally flat surfaces with boundaries is well understood and quite simple \cite{book.surfaces}: there is only the disk and cylinder. This results in a dramatic collapse of the topological expansion, with the only non-vanishing contributions given by $(g,n)=(0,1)$ and $(g,n)=(0,2)$. The partition functions on the disk and cylinder are then written as a path integral over the boundary modes appearing in the action (\ref{eq:IntroBounAct}). The second feature that allows us to determine (\ref{eq:IntroTopological}) exactly is that the path integral over the $\mathcal{N}=1$ boundary quantum mechanics is one-loop exact, due to the Duistermaat-Heckman theorem \cite{Duistermaat:1982vw,Stanford:2017thb}. Putting everything together, we derive the following expression for the partition function
\begin{equation}\label{eq:IntroPart}
Z(\beta) \simeq 
e^{S_0}
\frac{2\sqrt{2}}{\beta}\ ,
\qquad \qquad
Z(\beta_1,\beta_2) \simeq \frac{1}{\beta_1+\beta_2}\ ,
\qquad \qquad
Z(\beta_1,\dots,\beta_n) \simeq 0\ ,
\qquad n\ge 3\ ,
\end{equation}
where here $\beta_i$ is dimensionless, measured in units of $\gamma$. Since the Euclidean time coordinate $\tau$ is obtained from an analytic continuation of the retarded Bondi time $u$, this partition function is probing the spectrum of the Bondi Hamiltonian of the theory. The spectral density can be obtained from an inverse Laplace transform of $Z(\beta)$, and (quite surprisingly) one finds it is constant $\varrho(E)\simeq e^{S_0}2\sqrt{2}\Theta(E)$. This should be compared with ordinary CJ gravity, where the density is linear in the energy instead \cite{Godet:2020xpk}.

We continue in Section \ref{sec:3}, where the matrix model dual to $\mathcal{N}=1$ CJ supergravity is constructed and explored. Our first task is easy to state: we need to determine whether there is a random matrix model such that the connected ensemble average of a single trace matrix operator $\mathbb{O}(\beta)$ reproduces the partition function (\ref{eq:IntroPart}) to all orders in perturbation theory
\begin{equation}\label{eq:IntroMatching}
Z(\beta_1,\dots,\beta_n)\simeq \langle \mathbb{O}(\beta_1)\dots \mathbb{O}(\beta_n) \rangle_c\ .
\end{equation}
The solution to this problem is unique and remarkably simple. Consider an ensemble of $N$ dimensional squared Hermitian matrices $M$ with a Gaussian probability density measure $dM\,e^{-\frac{1}{2}N\,{\rm Tr}\,M^2}$. The ensemble average of the normalized eigenvalue density to leading order in $N\gg 1$ is the famous Wigner semi-circle distribution \cite{Wigner}
\begin{equation}\label{eq:Wigner}
\langle \rho(\lambda) \rangle = \frac{1}{2\pi} \sqrt{4-\lambda^2}+\mathcal{O}(1/N)\ .
\end{equation}
As understood in the nineties \cite{DiFrancesco:1993cyw}, the naive large $N$ is not enough to match with gravity. Apart from considering big matrices, we need to simultaneously rescale the eigenvalues $\lambda_i$ in the following way
\begin{equation}
\frac{1}{N}=\hbar \delta\ ,
\qquad \qquad
\lambda_i=\alpha_i\delta\ ,
\end{equation}
where $\delta\rightarrow 0$. The matrix model quantities $(N,\lambda_i)$ are replaced by the scaled parameters $(\hbar,\alpha_i)$. Using the matrix model loop equations (a set of recursion relations used for computing observables perturbatively) in Subsection \ref{sec:3.1} we prove the matching (\ref{eq:IntroMatching}) to all orders, subject to the following identifications\footnote{Here we have defined $\bar{M}=M/\delta$, so that $\bar{M}$ has the rescaled eigenvalues $\alpha_i$.}
\begin{equation}\label{eq:IntroOperator}
\mathbb{O}(\beta)=\int_{-\infty}^{+\infty}dp\,
{\rm Tr}\,e^{-\beta(\bar{M}^2+p^2)}\ ,
\qquad \qquad
\hbar=\frac{e^{-S_0}}{2\sqrt{2}}\ .
\end{equation}
This is the unique double scaled matrix model which ensures the matching in (\ref{eq:IntroMatching}). 

In agreement with other well understood holographic models in AdS, the operator $\mathbb{O}(\beta)$ takes the form ``~${\rm Tr}\,e^{-\beta H}$~", where in this case $H$ should be interpreted as the Bondi Hamiltonian. Interestingly, it contains two factorized contributions: a discrete matrix part $\bar{M}^2$ and a continuous non-relativistic free particle $p^2$. This peculiar structure (whose origin and significance is not fully understood) is exactly the same as the one obtained for ordinary CJ gravity in \cite{Kar:2022vqy,Kar:2022sdc}. It is therefore tempting to speculate this factorization is a feature of flat quantum gravity, where the continuous part might be somehow related to the infinite volume of flat space.

While fully non-perturbative effects are not under control in the metric description of $\mathcal{N}=1$ CJ supergravity, they are not particularly difficult to study using the matrix model. One can therefore use the holographic theory as a (non-unique) stable non-perturbative completion of flat quantum gravity. In practice, this simply means assuming the symbol $\simeq$ in (\ref{eq:IntroMatching}) can be replaced by an exact equality. All matrix model observables can be then computed exactly from the knowledge of the matrix model kernel $K(\alpha,\alpha')$ \cite{Eynard:2015aea}, which for our simple double scaled model is nothing more than the famous sine kernel
\begin{equation}\label{eq:IntroSine}
K(\alpha,\alpha')=\frac{1}{\pi}
\frac{\sin\left[(\alpha-\alpha')/\hbar\right]}{\alpha-\alpha'}\ .
\end{equation}
We should stress that in this context, the sine kernel is not an approximation, but instead the exact kernel of the double scaled model dual to $\mathcal{N}=1$ CJ supergravity. This provides us with unprecedented control over non-perturbative effects on a flat space quantum gravity theory. In Subsection \ref{sec:3.2} we use this to carefully study the fine grained spectrum of its Bondi Hamiltonian (Figure \ref{fig:2}), the late time behavior of the spectral form factor (Figure \ref{fig:3}), and ultra-low temperature dependence of the quenched free energy (Figure \ref{fig:4}). We also show the partition functions in (\ref{eq:IntroPart}) with $n\ge 3$ are not exactly zero, as they receive non-perturbative corrections (\ref{eq:105}). 

We finish in Section \ref{sec:4} with a brief discussion on promising future research directions one might pursue in order to push forward our understanding of these simple models of flat quantum gravity. Two appendices include some technical details regarding the $\mathcal{N}=1$ extension of the Maxwell algebra and the derivation of the matrix kernel (\ref{eq:IntroSine}) for the Gaussian model from first principles.

\section{Minimal CJ Supergravity}
\label{sec:2}

This section contains all gravitational calculations involving $\mathcal{N}=1$ CJ supergravity. We define its action from a BF gauge theory with an appropriate superalgebra, study its boundary degrees of freedom, and finally exactly compute the topological expansion of its Euclidean partition function.

\subsection{Formulation as a BF Gauge Theory}
\label{sec:2.1}

Following Cangemi and Jackiw \cite{Cangemi:1992bj}, in this subsection we construct the action of $\mathcal{N}=1$ CJ supergravity in first order formalism from a BF gauge theory. In this formulation there are two elementary fields: a space-time scalar $\mathbf{B}$ and a one-form connection $\mathbf{A}$, both valued on the algebra $\mathfrak{g}$ of the gauge group $G$. Under the group action, these fields transform as
\begin{equation}\label{eq:BFTrans}
\boldsymbol{B}\rightarrow G^{-1}\boldsymbol{B}G\ ,
\qquad \qquad
\boldsymbol{A}\rightarrow G^{-1}(d+\boldsymbol{A})G\ ,
\end{equation}
which means $\boldsymbol{B}$ is in the adjoint representation. While the gauge connection transforms with the usual anomalous term $G^{-1}dG$, its field strength $\boldsymbol{F}=d\boldsymbol{A}+\boldsymbol{A}\wedge \boldsymbol{A}$ also transforms in the adjoint. 

The construction of the BF action usually involves considering a gauge algebra which admits a bilinear form $\langle \cdot,\cdot \rangle:\mathfrak{g}\times \mathfrak{g}\rightarrow \mathbb{R}$ satisfying a number of properties.\footnote{Although we shall not consider them here, it is still possible to construct BF theories from algebras that do not admit a bilinear form with these properties~\cite{Grumiller:2020elf}.} For a superalgebra generated by $J_A\in \mathfrak{g}$, these properties can be stated as follows
\begin{equation}\label{eq:BilinearProp}
\begin{aligned}
{\rm Symmetric:}& \qquad
\langle J_A,J_B \rangle=(-1)^{|J_A|\cdot |J_B|}\langle J_B,J_A \rangle\ , \\[4pt]
{\rm Adjoint\,\,invariant}:& \qquad
\langle [J_A,J_C]_\pm,J_B \rangle=\langle J_A,[J_C,J_B]_\pm\rangle\ ,\\[4pt]
{\rm Non-degenerate}:& \qquad
\langle J_A,J_B \rangle=0\ , \qquad \forall\,\, J_A\in \mathfrak{g}
\qquad \Longleftrightarrow \qquad
J_B=0\  ,
\end{aligned}
\end{equation}
where $|J_A|=0,1$ for bosons and fermions respectively and $[J_A,J_B]_\pm$ a commutator or anti-commutator, where applicable. A canonical example of such a form is the Killing form of a semi-simple Lie superalgebra. The action of the BF gauge theory placed on a two-dimensional closed manifold $\mathcal{M}$ is
\begin{equation}\label{eq:BFAction}
I_{\rm BF}[\boldsymbol{A},\boldsymbol{B}]=
\int_{\mathcal{M}}
\langle \boldsymbol{B},\boldsymbol{F} \rangle\ ,
\end{equation}
which is immediately invariant under gauge transformations. Varying the action one finds its equations of motion
\begin{equation}\label{eq:BFEOM}
\delta I_{\rm BF}=0
\qquad \Longrightarrow \qquad
\boldsymbol{F}=0\ ,
\qquad
d\boldsymbol{B}+[\boldsymbol{A},\boldsymbol{B}]=0\ .
\end{equation}

To define a BF theory which has a geometric interpretation as a flat space dilaton gravity theory, one needs to pick a gauge algebra $\mathfrak{g}$ which contains as a subalgebra the Minkowski isometries, i.e. the Poincaré algebra, generated by two translations $P_a$ and a boost/rotation $J$, so that the gauge connection is expanded as $\boldsymbol{A}=e^aP_a+wJ+\cdots$. The one-form $e^a$ plays the role of the zweibein, related to the line element in the usual way $ds^2=g_{\mu \nu}dx^\mu dx^\nu=\eta_{ab}e^ae^b$, where Latin frame indices $a$ and $b$ are raised and lowered with the Minkowski/Euclidean metric $\eta_{ab}$. Apart from $e^a$, one has the spin connection $w_{ab}$, which in two dimensions is entirely determined by a single one-form component $w$ according to $w^{a}_{\,\,\,b}=\epsilon^a_{\,\,\,b} w$ with $\epsilon_{ab}$ the Levi-Civita symbol. In terms of these quantities, the torsion $T^a$ and curvature tensor $R_{ab}$ are written as
\begin{equation}
T^a=de^a+w\wedge \epsilon^a_{\,\,\,b} e^b\ ,
\qquad \qquad
R^a_{\,\,\,b}=\frac{1}{2}R^a_{\,\,\,bcd}e^c\wedge e^d=dw^a_{\,\,\,b}\ .
\end{equation}

There is however a problem with the above construction: the Poincaré algebra does not admit a non-degenerate bilinear form (\ref{eq:BilinearProp}). For this reason, Cangemi and Jackiw considered instead the minimal modification of Poincaré which allows for a bilinear form (\ref{eq:BilinearProp}), obtained by replacing the vanishing commutator between the two translations $P_a$ by a non-vanishing central element $I$. The resulting algebra is called the Maxwell algebra and has the following non-vanishing commutators
\begin{equation}\label{eq:MaxwellAlgebra}
[J,P_\pm] = \pm P_\pm\ ,
\qquad \qquad
[P_+,P_-]=I\ ,
\end{equation}
where $P_\pm$ are null translations.\footnote{The null generators are defined as ${\sqrt{2} P_\pm=P_1\pm i^{1-n_t}P_0}$, where $P_0$ and $P_1$ are the space and time components respectively and $n_t=0,1$ for the Euclidean and Lorentzian case.} The inclusion of the central element $I$ in order to avoid having a degenerate bilinear form is a procedure that works quite generally (see Section 4.1 in \cite{Grumiller:2020elf}). Writing the BF action (\ref{eq:BFAction}) using the Maxwell algebra (\ref{eq:MaxwellAlgebra}), one obtains the CJ gravity action.

The path for constructing the minimal CJ supergravity action is therefore quite clear. The first step is to enlarge the Maxwell algebra (\ref{eq:MaxwellAlgebra}) by introducing two fermionic generators $Q_\pm$. Requiring these generators have spin one-half $[J,Q_\pm]=\pm \frac{1}{2}Q_\pm$ and imposing the Jacobi identities, one arrives at the following $\mathcal{N}=1$ Maxwell superalgebra
\begin{equation}\label{eq:MaxwellSuperalgebra}
\begin{aligned}
{\rm Bosonic:}& \qquad
\hspace{15pt} [J,P_\pm]=\pm P_\pm\ ,
\qquad
\hspace{12pt} [P_+,P_-]=I\ , \\[4pt]
{\rm Mixed:}& \qquad
\hspace{15pt} [J,Q_\pm]=\pm\frac{1}{2}Q_\pm\ ,
\qquad 
[P_-,Q_+]=- \frac{1}{2} Q_-\ , \\[4pt]
{\rm Fermionic:}& \qquad
\lbrace Q_+,Q_+ \rbrace = P_+\ ,
\qquad
\hspace{12pt} \lbrace Q_+,Q_- \rbrace=I\ .
\end{aligned}
\end{equation}
This is almost the unique $\mathcal{N}=1$ extension of the Maxwell algebra in two-dimensions. There are two caveats. First, note there is an asymmetry between the relations satisfied by $Q_\pm$. More precisely, there are non-vanishing relations involving $Q_+$ with $P_-$ (second line) and with itself (third line) that have no corresponding relations for $Q_-$. One could have defined an analogous consistent superalgebra in which $Q_-$ has similar non-vanishing relations, instead of $Q_+$.\footnote{It is however not possible to define an $\mathcal{N}=1$ extension of the Maxwell algebra for which $\lbrace Q_\pm,Q_\pm \rbrace=\pm P_\pm$. Only one of these anti-commutators can be non-zero.} Secondly, another compatible superalgebra can be defined by setting $[P_-,Q_+]=\lbrace Q_+,Q_+\rbrace=0$ in (\ref{eq:MaxwellSuperalgebra}), which corresponds to the case previously studied in \cite{Soroka:2004fj}. Although we could have used this variant to define the CJ supergravity action, (\ref{eq:MaxwellSuperalgebra}) turns out being more convenient given that it leads to a better behaved asymptotic structure, related to the expected BMS symmetry of flat space. We provide more details on the $\mathcal{N}=1$ extensions of the Maxwell algebra in Appendix \ref{zapp:1}. In particular, we show how (\ref{eq:MaxwellSuperalgebra}) can be derived from an appropriate In\"{o}n\"{u}-Wigner contraction of the $\mathfrak{osp}(1|2)$ superalgebra, as well as provide an explicit six dimensional matrix representation. 

The non-trivial quadratic Casimir of (\ref{eq:MaxwellSuperalgebra}) is given by
\begin{equation}
C_2=\lbrace P_+,P_- \rbrace+\lbrace J,I \rbrace-\frac{1}{2}[Q_+,Q_-]\ .
\end{equation}
The matrix elements of the bilinear form $h_{AB}=\langle J_A,J_B \rangle$ can be obtained from the Casimir according to $C_2=h^{AB}J_{A}J_B$, which results in the following non-vanishing components
\begin{equation}\label{eq:Bilinear}
\langle P_+,P_- \rangle = 1\ ,
\qquad 
\langle J,I \rangle = 1\ ,
\qquad
\langle Q_+,Q_- \rangle =2\ ,
\end{equation}
satisfying all the properties listed in (\ref{eq:BilinearProp}).

We can now define the associated BF theory. The gauge connection $\boldsymbol{A}$ and scalar $\boldsymbol{B}$ have the following expansion in terms of the superalgebra generators\footnote{The fermionic components $\psi^\pm$ and $\lambda^\pm$ are Grassmann.}
\begin{equation}
\begin{aligned}
\boldsymbol{A}&=e^+P_++e^-P_-+wJ+AI+\psi^+Q_++\psi^-Q_-\ , \\[4pt]
\boldsymbol{B}&=x^+P_++x^-P_-+\Psi J+\Phi I+\lambda^+ Q_++\lambda^- Q_-\ ,
\end{aligned}
\end{equation}
so that the field strength $\boldsymbol{F}$ is given by
\begin{equation}
\begin{aligned}
\boldsymbol{F} & =
\Big[T^++\frac{1}{2}\psi^+\wedge\psi^+\Big]P_+
+T^- P_-
+dwJ
+\left[dA+e^+\wedge e^-+\psi^+\wedge \psi^-\right]I+\\[4pt]
& \hspace{10pt} +D\psi^+ Q_+
+\Big[
D\psi^-
-\frac{1}{2}e^-\wedge \psi^+\Big]Q_-\ ,
\end{aligned}
\end{equation}
where we have identified the components of the torsion $T^\pm=de^\pm \pm w\wedge e^\pm$ and the covariant derivative $D\psi^\pm=d\psi^\pm \pm \frac{1}{2}w\wedge \psi^\pm$.\footnote{Both in Lorentzian and Euclidean signature the flat metric in null coordinates is $\eta_{+-}=\eta_{-+}=1$, while the Levi-Civita symbol non-zero components are $\epsilon_{-+}=1$ and $\epsilon_{+-}=-1$.} Finally, the $\mathcal{N}=1$ CJ supergravity action, in first order formulation, is defined through the general BF action in (\ref{eq:BFAction}) 
\begin{equation}\label{eq:SCJAction}
\begin{aligned}
I_{\rm SCJ}&=
\int_{\mathcal{M}}
\bigg\lbrace
x^-
\Big[T^++\frac{1}{2}\psi^+\wedge\psi^+\Big]
+x^+T^-+\Phi dw
+\Psi
\left[dA+e^+\wedge e^-+\psi^+\wedge \psi^-\right]
+\\[4pt]
& \hspace{40pt}
+2\lambda^+
\Big[
D\psi^-
-\frac{1}{2}e^-\wedge \psi^+\Big]
-2\lambda^-D\psi^+
 \bigg\rbrace \ .
\end{aligned}
\end{equation}

To better understand this theory, let us momentarily turn off the fermions and examine its bosonic sector, which is ordinary CJ gravity. The scalars $x^\pm$ are Lagrange multipliers which enforce the zero torsion constraint $T^\pm=0$ on the space-time. It is therefore useful to momentarily switch to a second order description in terms of the metric $g_{\mu \nu}$, so that the bosonic action becomes
\begin{equation}\label{eq:CJAction}
I_{\rm SCJ}\big|_{{\rm fermions}=0}
=\frac{1}{2}\int_{\mathcal{M}}d^2x\sqrt{|g|}
\big[ \Phi R+2\Psi(\varepsilon^{\mu\nu}\partial_\mu A_\nu+1) \big]\ ,
\end{equation}
where we have defined $\sqrt{|g|}\varepsilon^{\mu \nu}=\epsilon^{\mu \nu}$. Crucially, the dilaton $\Phi$ appears linearly in the action, multiplying the Ricci scalar $R$, whose value in two space-time dimensions fully determines the Riemann tensor. As we see from (\ref{eq:SCJAction}), the term $\Phi R$ is actually not modified by the presence of the fermions. The other bosonic contributions involve the scalar $\Psi$, which also appears linearly in the action, coupled to a topological abelian gauge field $A_\mu$.

Going back to the full action (\ref{eq:SCJAction}), the fermionic dependence is finely tuned in order to ensure the theory is supersymmetric. The supersymmetry transformation, parametrized by the Grassmann function $\epsilon^\alpha$, appears in the BF formulation as a gauge transformation generated by a purely fermionic element of the superalgebra $\Theta=\epsilon^+ Q_++\epsilon^- Q_-$. Using the transformation of the fundamental fields of the BF theory (\ref{eq:BFTrans}), it is simple to deduce the following infinitesimal supersymmetry transformations for the individual components
\begin{equation}
\begin{aligned}
\delta_{\epsilon} e^+ &= \psi^+ \epsilon^+\ ,
\hspace{18.5pt}
\delta_{\epsilon} A =\psi^+\epsilon^-+\psi^-\epsilon^+\ ,
\hspace{19pt}
\delta_{\epsilon}\psi^+=D\epsilon^+\ ,
\hspace{27pt}
\delta_{\epsilon} \psi^-=D\epsilon^--\frac{1}{2} e^-\epsilon^+\ , \\[3pt]
\delta_{\epsilon} x^+&= \lambda^+\epsilon^+\ ,
\hspace{20pt}
\delta_{\epsilon} \Phi=\lambda^+\epsilon^-+\lambda^-\epsilon^+\ ,
\qquad
\delta_{\epsilon} \lambda^+=\frac{1}{2}\Psi \epsilon^+\ ,
\qquad
\delta_{\epsilon} \lambda^-=-\frac{1}{2}\Psi \epsilon^--\frac{1}{2} x^-\epsilon^+\ ,
\end{aligned}
\end{equation}
which, by construction, satisfy $\delta_{\epsilon}I_{\rm SCJ}=0$.

\subsubsection*{Solution to the Equations of Motion}

We now wish to construct a general class of solutions to the equations of motion of the CJ supergravity action. From the variation of the dilaton in (\ref{eq:CJAction}) or (\ref{eq:SCJAction}) we see all solutions must have a locally flat metric. Starting in Lorentzian signature, we fix the metric to the Bondi gauge and find the more general flat metric is parametrized by two arbitrary functions $T(u)$ and $P(u)$ of the retarded Bondi time $u$ according to
\begin{equation}\label{eq:LorBondi}
ds^2=-2
\big(P(u)r+T(u)\big)du^2
+2dudr\ .
\end{equation}
Since we are ultimately interested in the Euclidean partition function of this theory, we analytically continue the retarded time $u\rightarrow i\tau$. One might be worried by the fact the resulting metric becomes complex, essentially due to the non-diagonal components of the metric (\ref{eq:LorBondi}) in the Bondi gauge. We do not think this is problematic. Not only the associated path integral is finite and leads to a positive definite spectral density which can be non-perturbatively completed by a matrix model, but one can also check all the complex saddles used in these computations satisfy the criteria recently proposed in \cite{Kontsevich:2021dmb,Witten:2021nzp} to determine physically allowed complex manifolds.

For the analytically continued metric, a convenient choice for the frame fields $e^\pm$ which reproduces the correct metric is
\begin{equation}\label{eq:sol1}
e^+=i
\big(P(\tau)r+T(\tau)\big)d\tau-dr\ ,
\qquad
e^-=-id\tau\ ,
\qquad
w=-iP(\tau)d\tau\ ,
\end{equation}
where the spin connection $w$ (which trivially satisfies the flatness condition $dw\propto R =0$) is obtained from imposing the vanishing torsion $T^\pm=0$. Requiring the remaining components of $\boldsymbol{F}=0$ are satisfied, yields the following general solution for the abelian gauge field $A$ and the gravitino $\psi^\pm$
\begin{equation}\label{eq:sol2}
A=-i(r+N(\tau))d\tau\ ,
\qquad \qquad
\psi^\pm=-iH^\pm (\tau)d\tau\ ,
\end{equation}
where $N(\tau)$ and $H^\pm(\tau)$ are an ordinary and Grassmann functions respectively. 

From (\ref{eq:sol1}) and (\ref{eq:sol2}) one can write the general solution to the BF gauge connection $\boldsymbol{A}$. Same as in other similar two-dimensional theories \cite{Cardenas:2018krd,Afshar:2019tvp}, the $\tau$ and $r$ dependence of $\boldsymbol{A}$ admits the following factorized decomposition $\boldsymbol{A}=e^{rP_+}(d+\boldsymbol{a})e^{-rP_+}$ where $\boldsymbol{a}=\boldsymbol{a}_\tau(\tau)d\tau$ is given by\footnote{We thank Oscar Fuentealba and Hernán González for suggesting this parametrization of the solution.}
\begin{equation}\label{eq:60}
i\boldsymbol{a}=\left(
-T(\tau)P_++P_-+P(\tau)J
+N(\tau)I+H^+(\tau)Q_++H^-(\tau)Q_-
\right)d\tau\ .
\end{equation}
This description turns out being quite convenient for the analysis below. Although we could also analyze the solutions to the equations of motion for the scalar $\boldsymbol{B}$ in (\ref{eq:BFEOM}), this will not be necessary for the boundary conditions specified below.

\subsection{Asymptotic Boundary Conditions}
\label{sec:2.2}

So far we have considered the theory defined on a closed manifold $\mathcal{M}$. However, the interesting case arises when there is a boundary, as all the physical degrees of freedom localize on $\partial \mathcal{M}$. We assume there is an asymptotic circular boundary for large $r$, parametrized by the $\beta$-periodic coordinate $\tau\sim\tau+\beta$. To ensure the variation problem (\ref{eq:BFEOM}) of the BF theory is well defined, we add the following boundary term to the action
\begin{equation}\label{eq:BFAction2.0}
I_{\rm BF}[\boldsymbol{A},\boldsymbol{B}]=
\int_{\mathcal{M}}
\langle \boldsymbol{B},\boldsymbol{F} \rangle
-\frac{1}{2}\int_{\partial \mathcal{M}}
\langle \boldsymbol{B},\boldsymbol{A} \rangle
\ ,
\end{equation}
and relate the values of the fields according to $(\boldsymbol{B}+\gamma\boldsymbol{A}_\tau)\big|_{\partial \mathcal{M}}=0$ with $\gamma$ an arbitrary constant with units of inverse length.\footnote{In two-dimensional flat gravity there is no length scale that naturally arises in the definition of the theory, given that Newton's constant is dimensionless and there is no cosmological constant. One is therefore forced to introduce $\gamma$, an inverse length scale that all dimensionfull quantities in the theory are going to be measured with respect to.\label{foot:gamma}} The variation of the action leads to the same equations of motion as before~(\ref{eq:BFEOM}).

On top of the boundary condition relating $\boldsymbol{B}$ and $\boldsymbol{A}$ we want to further constraint the asymptotic fluctuations of the fields. Essentially, we impose a condition which, from the gravitational perspective, allows for a wiggly boundary but not more. Our guide for doing this is the solution to the equations of motion (\ref{eq:60}) which was constructed from a sensible and well behaved flat metric (\ref{eq:LorBondi}). We therefore require off-shell fluctuations of $\boldsymbol{A}$ have the following asymptotic behavior 
\begin{equation}\label{eq:AsymptoticCondition}
\boldsymbol{A}=e^{rP_+}(d+\boldsymbol{a})e^{-rP_+}+\mathcal{O}(1/r)\ ,
\end{equation}
where for $\boldsymbol{a}$ we pick
\begin{equation}\label{eq:Little_a}
i\boldsymbol{a}=(-T(\tau)P_++P_-+P(\tau)J+H(\tau)Q_+)d\tau\ ,
\end{equation}
parametrized by three functions. Compared to the general solution in (\ref{eq:60}), note that we are not allowing for fluctuations of $N(\tau)$ and $H^-(\tau)$ (we have also relabelled $H^+(\tau)\rightarrow H(\tau)$). This is an arbitrary choice made in order to ensure a convenient class of large gauge transformations and boundary action for the resulting theory.

With these boundary conditions in place, we can revisit the on-shell solutions to the equations of motion (\ref{eq:BFEOM}). For the gauge connection, imposing a flat connection $\boldsymbol{F}=0$ means the configuration parametrized by the functions (\ref{eq:Little_a}) is also valid in the interior of the manifold. Put simply, the on-shell solution for $\boldsymbol{A}$ is given by (\ref{eq:AsymptoticCondition}) without any $\mathcal{O}(1/r)$ corrections. For the scalar field one finds $\boldsymbol{B}=e^{rP_+}\boldsymbol{b}e^{-rP_+}$ where the boundary condition below (\ref{eq:BFAction2.0}) forces $\boldsymbol{b}=-\gamma\boldsymbol{a}_\tau$. The remaining equation of motion in (\ref{eq:BFEOM}) implies $d\boldsymbol{b}=-\gamma d\boldsymbol{a}_\tau=0$, meaning the three functions $(P(\tau),T(\tau),H(\tau))$ that parametrize the configurations are constant $(P_0,T_0,H_0)$ on-shell. The values $(P_0,T_0)$ determine a particular space-time metric, through the Euclidean version of (\ref{eq:LorBondi}), while $H_0$ sets the boundary value of the gravitino component $\psi^+$. In particular, the disk manifold with its center at $r=0$ is obtained from
\begin{equation}\label{eq:DiskSol}
{\rm Disk:} \qquad
(P_0,T_0,H_0)=\frac{2\pi}{\beta}(1,0,0)\ ,
\end{equation}
where the value of $P_0$ is fixed to avoid a conical singularity.

\subsubsection*{Large Gauge Transformations}

To better understand the symmetry structure of the theory as defined above, we now study the set of large gauge transformations allowed by our boundary conditions. These correspond to gauge transformations that do not vanish asymptotically, but instead preserve the form (\ref{eq:AsymptoticCondition}) while reshuffling the functions appearing in (\ref{eq:Little_a})
\begin{equation}\label{eq:0Trans}
(P(\tau),T(\tau),H(\tau))\quad \longrightarrow \quad
(\bar{P}(\tau),\bar{T}(\tau),\bar{H}(\tau))\ .
\end{equation}
From the gravitational perspective these are large diffeomorphisms that generate wiggles of the boundary. To study them, consider gauge transformations generated by ${\Theta=e^{rP_+}\theta e^{-rP_+}}$ with $\theta$ an element of the superalgebra that depends only on $\tau$. Requiring the behavior in (\ref{eq:AsymptoticCondition}) is preserved under gauge transformations gives the following condition
\begin{equation}\label{eq:56}
d\theta+[\boldsymbol{a},\theta]
=
\left[
\frac{\delta \bar{\boldsymbol{a}}}{\delta \bar{P}}\delta \bar{P}
+
\frac{\delta \bar{\boldsymbol{a}}}{\delta \bar{T}}\delta \bar{T}
+
\frac{\delta \bar{\boldsymbol{a}}}{\delta \bar{H}}\delta \bar{H}
\right]_{(\bar{P},\bar{T},\bar{H})=(P,T,H)}\ ,
\end{equation}
where $\bar{\boldsymbol{a}}$ is (\ref{eq:Little_a}) but with the transformed functions appearing on the right-hand side of (\ref{eq:0Trans}). Solving this constraint one finds large gauge transformations are parameterized by two ordinary functions $(\varepsilon(\tau),\sigma(\tau))$ and a Grassmann function $\eta(\tau)$ in the following way
\begin{equation}\label{eq:GaugeParameter}
\begin{aligned}
i\theta =& 
-\Big[\varepsilon(\tau)T(\tau)+\eta(\tau)H(\tau)+i\sigma'(\tau)\Big]P_+
+\varepsilon(\tau)P_-
+\Big[\varepsilon(\tau)P(\tau)-i\varepsilon'(\tau)\Big]J
-\sigma(\tau)I+\\[4pt]
&
+\Big[\varepsilon(\tau)H(\tau)-\eta(\tau)P(\tau)+2i\eta'(\tau)\Big]Q_+
+\eta(\tau)Q_-\ .
\end{aligned}
\end{equation}
The first order variations of the functions characterizing the asymptotic behavior are given by
\begin{equation}\label{eq:Variations}
\begin{aligned}
\delta P & =
\varepsilon(\tau) P'(\tau)+\varepsilon'(\tau)P(\tau)-i\varepsilon''(\tau) \ ,\\[4pt]
\delta T & = \varepsilon(\tau) T'(\tau)+2\varepsilon'(\tau)T(\tau)
+\eta(\tau) H'(\tau)+3\eta'(\tau)H(\tau)
+\sigma'(\tau) P(\tau)
+i\sigma''(\tau)\ , \\[4pt]
\delta H & = \varepsilon(\tau) H'(\tau)+\frac{3}{2}\varepsilon'(\tau)H(\tau)
-\eta(\tau) P'(\tau)
+\frac{1}{2}i\eta(\tau) P(\tau)^2
+2i\eta''(\tau)\ .
\end{aligned}
\end{equation}

Let us make a few observations about these important expressions. For the bosonic sector we recover the large gauge transformations of CJ gravity obtained in \cite{Afshar:2019axx}, which match with the coadjoint representation of the warped Virasoro algebra \cite{Afshar:2019tvp}. Note that if one sets $\sigma(\tau)$ and $\eta(\tau)$ to zero, the functions $(P(\tau),T(\tau),H(\tau))$ transform as fields of spin $s=1,2,\frac{3}{2}$ with respect to $\varepsilon(\tau)$. Finally, the transformation of $H(\tau)$ contains a non-linear contribution in the term $P(\tau)^2$, which means the associated superalgebra is non-linear. This is not unusual, as non-linear algebras arise in similar supersymmetric \cite{Cardenas:2018krd,Valcarcel:2018kwd} and higher spin theories \cite{Gonzalez:2018enk,Afshar:2020dth}.

To explicitly figure out the superalgebra associated to (\ref{eq:Variations}), let use the covariant phase space formalism to compute the charges that generate the transformations. The variation of the bulk term in the BF action (\ref{eq:BFAction2.0}) gives the pre-symplectic potential that we vary to obtain the pre-symplectic form $\widetilde{\Omega}(\delta_1,\delta_2)=\langle \delta_1\boldsymbol{B},\delta_2 \boldsymbol{A}\rangle-
\langle \delta_2\boldsymbol{B},\delta_1 \boldsymbol{A}\rangle$. Keeping the first variation $\delta_1$ arbitrary while fixing the second $\delta_2=\delta_{\rm gauge}$ to the gauge transformation in (\ref{eq:GaugeParameter}), one gets
\begin{equation}
\widetilde{\Omega}(\delta,\delta_{\rm gauge}) = d\langle \delta \boldsymbol{B},\Theta \rangle
-\langle \delta(d\boldsymbol{B}+[\boldsymbol{A},\boldsymbol{B}]),\Theta \rangle\ .
\end{equation}
Since the second term vanishes on-shell (\ref{eq:BFEOM}), the resulting expression is an exact form, meaning the variation of the charge that generates the gauge transformation is
\begin{equation}\label{eq:ChargeVariation}
\delta \mathcal{Q}[\theta]=
\int_0^\beta  d\tau\,
\langle \delta \boldsymbol{B},\Theta \rangle\big|_{\partial \mathcal{M}}=
-\gamma
\int_0^\beta  d\tau\,
\langle \delta \boldsymbol{a}_\tau,\theta \rangle\ ,
\end{equation}
where in the second equality we used the boundary condition below (\ref{eq:BFAction2.0}). Note there is an unusual feature in the way we have defined the charge, given that in general one should not integrate over $\tau$, but instead evaluate at a fixed time. However, defining the charges in this way has been shown to be more adequate in setups similar to this one \cite{Cadoni:2000ah,Afshar:2015wjm,Grumiller:2017qao}. Using the expressions above we can evaluate (\ref{eq:ChargeVariation}) and arrive at the final expression for the charges
\begin{equation}\label{eq:Charges}
\mathcal{Q}[\theta]=
\gamma^2
\int_0^\beta d\tau
\left(
\varepsilon(\tau) T(\tau)
+\sigma(\tau) P(\tau)
+2 \eta(\tau) H(\tau)
\right)\ ,
\end{equation}
where we have conveniently added an additional factor of $-\gamma$ in their definition to have $\mathcal{Q}[\theta]$ dimensionless. The charges are integrable, given that the bilinear form only picks up the components of $\theta$ that are independent of the functions $(P(\tau),T(\tau),H(\tau))$.

The superalgebra satisfied by these charges with respect to the Poisson brackets is obtained from $\lbrace \mathcal{Q}[\theta_1],\mathcal{Q}[\theta_2] \rbrace_{\rm PB}=\delta_{\theta_2}Q[\theta_1]$. Computing the variation on the right-hand side of this expression is a straightforward exercise which gives
\begin{equation}\label{eq:86}
\begin{aligned}
\lbrace \mathcal{Q}[\theta_1],\mathcal{Q}[\theta_2] \rbrace_{\rm PB} & =
\mathcal{Q}[\bar{\varepsilon},\bar{\sigma},\bar{\eta}]
+i\gamma^2
\int_0^\beta d\tau
\eta_1\eta_2P^2
-i\gamma^2
\int_0^\beta d\tau
(\varepsilon_1'\sigma_2'
-\varepsilon_2'\sigma_1'
+4\eta_1'\eta_2')
\ ,
\end{aligned}
\end{equation}
where the functions appearing on the first term on the right-hand side are
\begin{equation}\label{eq:65}
\begin{aligned}
\bar{\varepsilon}(\tau) & =\varepsilon_1(\tau)\varepsilon_2'(\tau)-\varepsilon_2(\tau)\varepsilon_1'(\tau)\ ,\\[4pt]
\qquad
\bar{\sigma}(\tau) & =\varepsilon_1(\tau)\sigma_2'(\tau)-\varepsilon_2(\tau)\sigma_1'(\tau)
+2\big(\eta_1(\tau)\eta_2(\tau)\big)'\ ,\\[4pt]
\bar{\eta}(\tau) & = 
\big(\varepsilon_1(\tau)\eta_2'(\tau)-\varepsilon_2(\tau)\eta_1'(\tau)\big)
-
\frac{1}{2}\big(\varepsilon_1'(\tau)\eta_2(\tau)-\varepsilon_2'(\tau)\eta_1(\tau)\big)
\ .
\end{aligned}
\end{equation}
The second term in (\ref{eq:86}) cannot be written in terms of the charges, as it contains the non-linear contribution in $P(\tau)^2$. Finally, the third term is independent of the functions $(P(\tau),T(\tau),H(\tau))$ and therefore corresponds to a central extension. 

To get a better hold of the superalgebra, it is convenient to perform the following Fourier mode decomposition of the generators
\begin{equation}
L_n=\mathcal{Q}\left[\frac{\beta}{2\pi} e^{in\frac{2\pi}{\beta} \tau},0,0\right]\ ,
\quad
J_n=
\mathcal{Q}\left[0,\frac{1}{2\pi \gamma^2}e^{in\frac{2\pi}{\beta} \tau},0\right]\ ,
\quad
G_n=\mathcal{Q}\left[0,0,i\frac{\sqrt{\beta}}{2\pi \gamma} e^{in\frac{2\pi}{\beta} \tau}\right]\ .
\end{equation}
Replacing Poisson brackets by (anti-)commutators in the usual way, one finds
\begin{equation}\label{eq:AsymptoticAlgebra1}
\begin{aligned}
\relax [ L_n, L_m ] & =
(n-m)  L_{n+m}\ , \\[4pt]
[ L_n,J_m ] & = 
-mJ_{n+m}
+n^2\delta_{n+m,0}
\ , \\[4pt]
[ L_n,G_r ] & =
\left(\frac{n}{2}-r\right)G_{n+r}\ ,\\[4pt]
\lbrace G_r,G_s \rbrace & = 
2(r+s)J_{r+s}
+\sum_{q\in \mathbb{Z}}
J_qJ_{(r+s)-q}
-4 r^2\delta_{r+s,0}
\ .
\end{aligned}
\end{equation}
The bosonic sector (first two lines) is nothing more than a particular central extension of the warped Virasoro algebra \cite{Afshar:2015wjm}. The addition of the fermionic generator $G_r$ gives its supersymmetric extension. The non-linearity of the superalgebra can be hidden by exchanging $J_n$ with the following twisted Sugawara generator
\begin{equation}
M_n=2nJ_n
+\sum_{q\in \mathbb{Z}}
J_qJ_{n-q}\ .
\end{equation}
In terms of $M_n$, the superalgebra is linear and closes to
\begin{equation}\label{eq:sBMS3}
\begin{aligned}
\relax [ L_n, L_m ] & =
(n-m)  L_{n+m}\ , \\[4pt]
[ L_n,M_m ] & = 
(n-m)M_{n+m}
-2 n^3\delta_{n+m,0}
\ , \\[4pt]
[ L_n,G_r ] & =
\left(\frac{n}{2}-r\right)G_{n+r}\ ,\\[4pt]
\lbrace G_r,G_s \rbrace & = 
M_{r+s}
-4r^2\delta_{r+s,0}
\ ,
\end{aligned}
\end{equation}
that is nothing more than the three dimensional BMS superalgebra \cite{Barnich:2014cwa}, with $L_n$ and $M_n$ playing the role of superrotations and supertranslations respectively.

\subsection{Boundary Dynamics}
\label{sec:2.3}

To properly characterize the boundary degrees of freedom of the theory, we would like to derive the effective action that controls its dynamics. This is easy to do for large values of $r$, by evaluating the BF action (\ref{eq:BFAction2.0}) using the boundary conditions $(\boldsymbol{B}+\gamma\boldsymbol{A}_\tau)\big|_{\partial \mathcal{M}}=0$ and (\ref{eq:AsymptoticCondition}), which gives
\begin{equation}\label{eq:BdyAction1}
I_\partial=
\frac{\gamma}{2}\int_0^\beta d\tau\,
\langle \boldsymbol{a}_\tau,\boldsymbol{a}_\tau \rangle+\mathcal{O}(1/r)
=\gamma\int_0^\beta d\tau\, 
T(\tau)+\mathcal{O}(1/r)\ .
\end{equation}
This simple result for the boundary action is one of the reasons we picked the boundary conditions in (\ref{eq:Little_a}). Note $I_\partial$ coincides with the charge (\ref{eq:Charges}) associated to a large gauge transformation with constant $\varepsilon(\tau)$ and vanishing $\sigma(\tau)=\eta(\tau)=0$. This observation will be important when computing the Euclidean path integral.

It is useful to derive a different form for (\ref{eq:BdyAction1}), obtained by acting with large gauge transformations on an on-shell solution $(P_0,T_0,H_0)$. To do so, we follow \cite{Cardenas:2018krd} and note a flat connection $\boldsymbol{a}$ can be written as $\boldsymbol{a}=g^{-1}dg$ with $g=g(\tau)$ an element of the Maxwell supergroup. Equating $\boldsymbol{a}=g^{-1}dg$ to the expression of $\boldsymbol{a}$ in terms of the functions $(P(\tau),T(\tau),H(\tau))$ given in (\ref{eq:Little_a}), one finds the more general expression for $g(\tau)$ in an Euler-Gauss decomposition is
\begin{equation}\label{eq:Little_g}
g(\tau)=e^{F(\tau)P_-+\vartheta(\tau)Q_-+h(\tau)I}
e^{-\ln [iF'(\tau)] J}
e^{ih'(\tau)P_++\frac{2\vartheta'(\tau)}{\sqrt{-iF'(\tau)}} Q_+}
\ .
\end{equation}
The supergroup element is parametrized by two arbitrary functions $(F(\tau),h(\tau))$ and the Grassmann function $\vartheta(\tau)$, which are related to the boundary modes $(P(\tau),T(\tau),H(\tau))$ in the following way
\begin{equation}\label{eq:Relations}
\begin{aligned}
P(\tau) & =-i\frac{F''(\tau)}{F'(\tau)}\ , \\[4pt]
T(\tau) & =\frac{-1}{F'(\tau)}\left[ 
h'(\tau)F''(\tau)+2\vartheta'(\tau)\vartheta''(\tau)-F'(\tau)h''(\tau)
\right]\ , \\[4pt]
H(\tau) & = 2
\frac{F'(\tau)\vartheta''(\tau)-F''(\tau)\vartheta'(\tau)}{[-iF'(\tau)]^{3/2}}\ .
\end{aligned}
\end{equation}
This gives a more explicit expression of the boundary action in (\ref{eq:BdyAction1})
\begin{equation}\label{eq:BdyAction2}
I_\partial[F,h,\vartheta]=-\gamma\int_0^\beta  
\frac{d\tau}{F'(\tau)}\left[ 
h'(\tau)F''(\tau)+2\vartheta'(\tau)\vartheta''(\tau)
\right]\ ,
\end{equation}
where we have dropped a boundary term and corrections that vanish when $r\rightarrow \infty$.

The three modes $(F(\tau),h(\tau),\vartheta(\tau))$ control the boundary degrees of freedom. They parametrize large gauge transformations acting on an on-shell configuration $(P_0,T_0,H_0)$. Although not evident from our derivation, the action (\ref{eq:BdyAction2}) is not completely general, but only corresponds to configurations obtained from solutions with fixed $P_0$ and vanishing $T_0=H_0=0$. To see this, we need to translate between $(F(\tau),h(\tau),\vartheta(\tau))$ and the infinitesimal description of the transformation (\ref{eq:GaugeParameter}) in terms of $(\varepsilon(\tau),\sigma(\tau),\eta(\tau))$. It turns out these fields are related in the following way (see also \cite{Cardenas:2018krd})
\begin{equation}
F(\tau)=e^{i P_0(\tau+\varepsilon(\tau))}\ ,
\qquad \qquad
h(\tau)=i\sigma(\tau)\ ,
\qquad \qquad
\vartheta(\tau)=\sqrt{-iF'(\tau)}
\eta(\tau)\ .
\end{equation}
Using these relations we can expand the action (\ref{eq:BdyAction2}) and match with the expansion of~(\ref{eq:BdyAction1}) using~(\ref{eq:Variations})
\begin{equation}\label{eq:82}
I_\partial[\varepsilon,\sigma,\eta]=\gamma\int_0^\beta d\tau\,T(\tau)=\gamma\int_0^\beta d\tau
\Big[ 
T_0+\delta T+\frac{1}{2}\delta^2T+\dots
\Big]\ ,
\end{equation}
with $(P_0,T_0,H_0)=(P_0,0,0)$. Setting $P_0=\frac{2\pi}{\beta}$ the action (\ref{eq:BdyAction2}) controls the dynamics of the boundary modes around the disk solution (\ref{eq:DiskSol}), with $F(\tau)$ and $h(\tau)$ periodic functions. Although not necessary for our purposes, it would be interesting to derive the boundary action (\ref{eq:BdyAction2}) but for arbitrary values of $(P_0,T_0,H_0)$, as done in \cite{Afshar:2019axx} for bosonic CJ gravity. 

Let us now study some of the features of the boundary action (\ref{eq:BdyAction2}). From its variation, one finds the following equations of motion
\begin{equation}\label{eq:bdyEOM}
\frac{d}{d\tau}\left[\frac{F''(\tau)}{F'(\tau)}\right]=0\ ,
\qquad
\frac{d}{d\tau}\left[ \frac{h''(\tau)}{F'(\tau)}+2\frac{\vartheta'(\tau)\vartheta''(\tau)}{F'(\tau)^2} \right]=0\ ,
\qquad
\frac{d}{d\tau}\left[\left(\frac{\vartheta'(\tau)}{F'(\tau)}\right)'+\frac{\vartheta''(\tau)}{F'(\tau)}\right]=0\ .
\end{equation}
Using (\ref{eq:Relations}) these equations can be shown to be equivalent to $P'(\tau)=T'(\tau)=H'(\tau)=0$, which are nothing more than the bulk equations of motion (\ref{eq:BFEOM}), as discussed above (\ref{eq:DiskSol}). There are four bosonic and two fermionic independent symmetry transformations that preserve the equations of motion, given by
\begin{equation}
\begin{aligned}
F(\tau)\quad &\longrightarrow \quad
c_1F(\tau)+c_2\ , \\[4pt]
h(\tau) \quad &\longrightarrow \quad
h(\tau)+c_3F(\tau)+c_4+\epsilon_2\vartheta(\tau)\ ,\\[4pt]
\vartheta(\tau) \quad &\longrightarrow \quad
\sqrt{c_1}\Big(\vartheta(\tau)+\epsilon_1+\frac{1}{2}\epsilon_2F(\tau)\Big)\ ,
\end{aligned}
\end{equation}
where $c_i$ and $\epsilon_i$ are ordinary and Grassmann parameters respectively. In particular note there is a single transformation, the one controlled by $\epsilon_2$, which mixes the bosonic and fermionic fields. This means the boundary theory has $\mathcal{N}=1$ supersymmetry, reflecting the same amount of supersymmetry of the parent CJ supergravity action (\ref{eq:SCJAction}).

\subsection{Euclidean Partition Function}
\label{sec:2.4}

We can now finally turn our attention to the observable of CJ supergravity we are mostly interested in: the Euclidean partition function. Formally, it is defined through the following path integral
\begin{equation}\label{eq:PartFunc}
Z(\beta_1,\dots,\beta_n)=\int \mathcal{D}X\,e^{-I_{\rm SCJ}[X]+S_0\chi(\mathcal{M})}\ ,
\end{equation}
where the action is given in (\ref{eq:SCJAction}) and $\mathcal{D}X$ is the integral measure over all the fields in the theory.\footnote{Apart from the bulk contribution to the CJ supergravity action in (\ref{eq:SCJAction}), one must also include the appropriate boundary term that ensures the variational problem is well defined. While in the BF formulation given through (\ref{eq:BFAction2.0}) the boundary term is explicit, it would be interesting to work it out directly in the gravitational formulation. For ordinary CJ gravity this was done in Section 2.2 of \cite{Godet:2021cdl}.} We added to the exponent a term proportional to the Euler Characteristic $\chi(\mathcal{M})$ of the manifold $\mathcal{M}$, controlled by the parameter $S_0\in \mathbb{R}_+$. Since this is a topological invariant, it can be included in the definition of the theory without modifying any of the analysis in the previous subsections.

For (\ref{eq:PartFunc}) to make sense, we need to specify the boundary conditions. We allow for $n$-asymptotic boundaries, each of them locally parametrized by a $\beta_i$-periodic coordinate $\tau_i$, defined from the analytic continuation indicated below equation (\ref{eq:LorBondi}). For each of these boundaries, we constraint to off-shell configurations with the behavior given in (\ref{eq:AsymptoticCondition}). While for the bosonic fields we consider periodic boundary conditions when going around the boundary circle, for the fermionic fields we take anti-periodic (Neveu-Schwarz). The path integral not only includes a sum over bulk geometries consistent with these boundary conditions, but also a summation over inequivalent bulk spin structures.

Computing the path integral in (\ref{eq:PartFunc}) is very challenging. One can make progress by using that all two-dimensional orientable manifolds are classified by their genus $g$ and number of boundaries~$n$. Using the Euler characteristic is given by $\chi(\mathcal{M})=2(1-g)-n$, one arrives at the following topological expansion
\begin{equation}\label{eq:TopExp2}
Z(\beta_1,\dots,\beta_n)\simeq \sum_{g=0}^{\infty}(e^{-S_0})^{2(g-1)+n}
Z_g(\beta_1,\dots,\beta_n)\ .
\end{equation}
The symbol $\simeq$ reminds us this is nothing more than a series expansion, i.e. it is only equal to the actual partition function (\ref{eq:PartFunc}) up to corrections of order $\mathcal{O}(e^{-e^{S_0}})$. Each of the terms in the expansion are determined by $Z_g(\beta_1,\dots,\beta_n)$, which are defined in the same way as (\ref{eq:PartFunc}) but with the important difference we only include contributions from manifolds of fixed genus $g$.

It is now that the details of CJ supergravity become important, specifically the linear dependence of the dilaton in the action (\ref{eq:SCJAction}). Taking the integration contour over $\Phi$ along a purely imaginary line allows us to trivially solve the path integral of the dilaton and obtain a Dirac delta $\delta(R)$. Since in two-dimensions the Riemann tensor is fully determined by the Ricci scalar
\begin{equation}
R_{\mu \nu\rho \sigma}=\frac{R}{2}
(g_{\mu \rho}g_{\nu \sigma}
-g_{\mu \sigma}g_{\nu \rho})\ ,
\end{equation}
the Dirac delta effectively becomes $\delta(R_{\mu \nu \rho \sigma})$. This highly constraints the integral over metrics, as instead of having arbitrary metric fluctuations, one only needs to consider locally flat manifolds. The classification of two-dimensional orientable flat surfaces with asymptotic boundaries is extremely simple~\cite{book.surfaces}: there is only the disk and cylinder. Putting everything together, the integral over the dilaton implies a spectacular cancellation of most terms in the topological expansion, so that one is simply left with\footnote{One might wonder about a disk or cylinder with an arbitrary number of circular boundaries in their interior. These do not contribute since we are constraining ourselves to asymptotic boundaries, i.e. boundaries for which the distance of any bulk point to the boundary is infinite. These other surfaces should be considered in the finite cut-off version of the theory.}
\begin{equation}\label{eq:DiskCylTopExp}
\begin{aligned}
Z(\beta)& \simeq e^{S_0}Z_{\rm disk}(\beta)\ , \\[4pt]
Z(\beta_1,\beta_2) & \simeq Z_{\rm cylinder}(\beta_1,\beta_2)\ , \\[4pt]
Z(\beta_1,\dots,\beta_n) & \simeq 0\ ,
\end{aligned}
\end{equation}
where $n\ge 3$. While the same reduction occurs for ordinary CJ gravity, a different mechanism also results in the same effect for certain supersymmetric extensions of JT gravity and deformations thereof~\cite{Stanford:2019vob,Rosso:2021orf}. 

All we have to do, is compute the disk and cylinder partition functions. To do so, it is convenient to write these quantities using the BF formulation of the theory, whose partition function is given by
\begin{equation}\label{eq:BFPart}
Z_{\rm BF}=\int \mathcal{D}\boldsymbol{A}\mathcal{D}\boldsymbol{B}e^{-I_{\rm BF}[\boldsymbol{A},\boldsymbol{B}]}=
\int \mathcal{D}\boldsymbol{A}
\delta(\boldsymbol{F})
e^{-\frac{\gamma}{2}\int_{\partial \mathcal{M}}
\langle \boldsymbol{A},\boldsymbol{A} \rangle}\ ,
\end{equation}
where, similarly as in the gravitational description, the integral over $\boldsymbol{B}$ localizes the remaining integral over flat connections $\boldsymbol{F}=0$. Path integrals of this kind were studied long ago in \cite{Witten:1991we}. Performing the standard gauge fixing via the Fadeed-Popov method, it was shown that the resulting measure of the path integral is obtained from the Pfaffian of the following symplectic form in the space of flat connections
\begin{equation}\label{eq:SympForm}
\Omega(\delta_1\boldsymbol{A},\delta_2\boldsymbol{A})=\gamma^2 c_0  \int_{\mathcal{M}}\langle \delta_1\boldsymbol{A}\wedge \delta_2\boldsymbol{A}  \rangle\ ,
\end{equation}
where $c_0$ is an arbitrary dimensionless constant and we are omitting the comma in the bilinear form. The one-forms in the space of flat connections are given by $\delta_i \boldsymbol{A}$, which are variations that preserve $\boldsymbol{F}=0$ to first order. The path integral is weighted by the boundary action (\ref{eq:BdyAction1}), corresponding to an $\mathcal{N}=1$ supersymmetric quantum mechanics.

Exactly computing the path integral of a quantum mechanical system is still, in general, a challenging task. Usually, one is instead able to perform a perturbative loop expansion. The Duistermaat-Heckman theorem \cite{Duistermaat:1982vw} singles out certain situations in which the simple one-loop computation is not only an approximation but it actually coincides with the exact answer. As explained in \cite{Stanford:2017thb}, there are two conditions that must be satisfied for the theorem to apply: the integration space must be symplectic and the action weighting the integral must generate a $U(1)$ symmetry of the manifold via the Poisson brackets.

The first condition is automatically satisfied by the BF partition function, given that the integral is over the symplectic manifold of flat connections (\ref{eq:SympForm}). For the second requirement, note that using the boundary condition (\ref{eq:AsymptoticCondition}) we can write the boundary action appearing in (\ref{eq:BFPart}) as 
\begin{equation}
I_\partial=\gamma \int_0^\beta d\tau\,T(\tau)=\mathcal{Q}_\tau\ ,
\end{equation}
where $\mathcal{Q}_\tau$ is a particular generator (\ref{eq:Charges}) of a large gauge transformation 
\begin{equation}
\mathcal{Q}_\tau\equiv\mathcal{Q}[\varepsilon,\sigma,\eta]\ ,
\qquad
{\rm with}
\qquad
\varepsilon(\tau)=\frac{1}{\gamma}\ ,
\qquad
\sigma(\tau)=\eta(\tau)=0\ .
\end{equation}
The action of this charge via the Poisson brackets on an arbitrary flat connection characterized by the functions $(P(\tau),T(\tau),H(\tau))$ can be worked out from the infinitesimal variations in (\ref{eq:Variations})
\begin{equation}
\lbrace S(\tau),\mathcal{Q}_\tau \rbrace_{\rm PB}=\delta_{(\varepsilon,\sigma,\eta)} S(\tau)=\frac{1}{\gamma}S'(\tau)\ ,
\qquad {\rm where} \qquad
S(\tau)=(P(\tau),T(\tau),H(\tau))\ .
\end{equation}
This shows the boundary action in the path integral is indeed generating $\tau$ transations around the circle. Altogether, the Duistermaat-Heckman theorem applies and the BF path integral (\ref{eq:BFPart}), which determines the disk and cylinder partition functions in (\ref{eq:DiskCylTopExp}), can be calculated exactly from a simple one-loop computation. 

Before performing this calculation, we need to comment on an important subtlety regarding the symplectic space of flat gauge connections. Although one of the defining properties of a symplectic form is its non-degeneracy, this is actually not immediately satisfied by (\ref{eq:SympForm}) when evaluated on the asymptotic boundary. To see this, we use that since the variations $\delta_i\boldsymbol{A}$ correspond to large gauge transformations generated by $\Theta_i$ as $\delta_i\boldsymbol{A}=d\Theta_i+[\boldsymbol{A},\Theta_i]$, equation (\ref{eq:SympForm}) can be reduced to a boundary integral
\begin{equation}
\Omega(\delta_1\boldsymbol{A},\delta_2\boldsymbol{A})=\gamma^2 c_0  \int_0^\beta d\tau \langle \theta_1 , \delta_2\boldsymbol{a}_\tau \rangle\ ,
\end{equation}
where we have used $(d\Theta_1+[\Theta_1,\boldsymbol{A}])\wedge \delta_2\boldsymbol{A}=d(\Theta_1\wedge \delta_2\boldsymbol{A})$ for variations that linearly preserve $\boldsymbol{F}=0$. Using (\ref{eq:GaugeParameter}) and (\ref{eq:Variations}) this can be evaluated and written as
\begin{equation}
\begin{aligned}
\Omega(\delta_1\boldsymbol{A},\delta_2\boldsymbol{A})=& \gamma^2 c_0  \int_0^\beta d\tau \Big\lbrace
\bar{\varepsilon}(\tau)T(\tau)
+\big[\bar{\sigma}(\tau)+i\eta_1(\tau)\eta_2(\tau)P(\tau)\big]P(\tau)+2\bar{\eta}(\tau)H(\tau)+\\[4pt]
&\hspace{130pt}
-i\big(\varepsilon_1'(\tau)\sigma_2'(\tau)-\varepsilon_2'(\tau)\sigma_1'(\tau)+4\eta_1'(\tau)\eta_2'(\tau)\big)
\Big\rbrace
\ ,
\end{aligned}
\end{equation}
where the barred functions are defined in (\ref{eq:65}).\footnote{This expression generalizes the symplectic form derived in \cite{Kar:2022sdc} for bosonic CJ gravity.} For both the disk and cylinder we need to evaluate this for on-shell solutions with $(P(\tau),T(\tau),H(\tau))=(P_0,T_0,0)$, which gives
\begin{equation}\label{eq:80}
\begin{aligned}
\Omega=& \gamma^2 c_0  \int_0^\beta d\tau \Big\lbrace
T_0 \delta \varepsilon \wedge \delta \varepsilon'
+P_0\delta \varepsilon \wedge \delta \sigma'
-i\delta \varepsilon'\wedge \delta \sigma'
-2i\left(
\delta \eta'\wedge \delta \eta'
-(P_0/2)^2\delta \eta \wedge \delta \eta \right)
\Big\rbrace
\ ,
\end{aligned}
\end{equation}
where we have written $\Omega$ using form notation. This expression becomes more transparent when expanding the functions in their respective Fourier series
\begin{equation}\label{eq:Fourier}
\varepsilon(\tau)=\sum_{n\in \mathbb{Z}}\varepsilon_n e^{in(\frac{2\pi}{\beta})\tau}\ ,
\qquad
\sigma(\tau)=\sum_{n\in \mathbb{Z}}\sigma_n e^{in(\frac{2\pi}{\beta})\tau}\ ,
\qquad
\eta(\tau)=\sum_{n\in \mathbb{Z}+\frac{1}{2}}
\eta_n e^{in(\frac{2\pi}{\beta})\tau}\ ,
\end{equation}
where for $\eta(\tau)$ we have half-integer modes since we impose Neveu-Schwarz boundary conditions $\eta(\tau+\beta)=-\eta(\tau)$. The reality condition on these functions constraints $\varepsilon^\ast_n=\varepsilon_{-n}$ (and similarly for the other coefficients), so that the symplectic manifold is parametrized by the complex parameters $(\varepsilon_n,\sigma_n,\eta_n)$ with $n\ge 0$. In this parametrization, (\ref{eq:80}) becomes
\begin{equation}\label{eq:81}
\begin{aligned}
\Omega=&-2\pi i \gamma^2 c_0
\sum_{n\ge 0}n
\bigg[
2T_0\delta \varepsilon_n \wedge \delta \varepsilon_n^\ast
+
\Big(
P_0+\frac{2\pi n}{\beta}
\Big)\delta \varepsilon_n\wedge \delta \sigma_n^\ast
+\Big(
P_0-\frac{2\pi n}{\beta}
\Big)
\delta \sigma_n \wedge \delta \varepsilon_n^\ast
\bigg]+ \\[4pt]
&-4\beta i\gamma^2 c_0
\sum_{n> 0}
\left[
\left(\frac{2\pi n}{\beta}\right)^2
-\left(\frac{P_0}{2}\right)^2\right]
\delta\eta_n\wedge \delta \eta_n^\ast
\ .
\end{aligned}
\end{equation}
From this expression, we see $\Omega$ is always degenerate along the $(\varepsilon_0,\sigma_0)$ directions. This means the actual symplectic space over which the path integral is performed corresponds to $(\varepsilon_n,\sigma_n,\eta_n)$ after modding out all degenerate directions (or zero modes), which may be enhanced for special values of~$P_0$. As we shall see, these zero modes completely determine the $\beta$ scaling of the one-loop determinant according to
\begin{equation}\label{eq:ZeroModes}
Z_{\rm one-loop}(\beta)\propto \beta^{\frac{1}{2}(n_f-n_b)}\ ,
\end{equation}
where $n_b$ and $n_f$ are the number of bosonic and fermionic zero modes.

\subsubsection*{Disk Partition Function}

Let us now compute the disk partition function from its one-loop computation
\begin{equation}\label{eq:83}
Z_{\rm disk}(\beta)=
e^{-I_\partial^{\rm (on-shell)}}
\int 
\mathcal{D}\varepsilon(\tau)
\mathcal{D}\sigma(\tau)
\mathcal{D}\eta(\tau)
{\rm Pf}(\Omega)
e^{-I_\partial^{(2)}[\varepsilon(\tau),\sigma(\tau),\eta(\tau)]}\ ,
\end{equation}
where we are implicitly removing the degenerate directions in the measure. From (\ref{eq:BdyAction1}) the on-shell boundary action is given by $I_{\partial}^{\rm (on-shell)}=\gamma \beta T_0 $, which vanishes since for the disk solution~(\ref{eq:DiskSol}) we have $T_0=0$. The quadratic action $I_\partial^{(2)}[\varepsilon,\sigma,\eta]$ around an on-shell configuration with arbitrary $(P_0,T_0)$ and $H_0=0$ is obtained from (\ref{eq:82}) and (\ref{eq:Variations})
\begin{equation}\label{eq:Quad}
\begin{aligned}
I_\partial^{(2)} & =\gamma\int_0^\beta d\tau
\Big[
T_0\varepsilon'(\tau)^2
+\varepsilon'(\tau)\big(P_0\sigma'(\tau)
+i\sigma''(\tau)\big)
+2i\left(\eta'(\tau)\eta''(\tau)-(P_0/2)^2\eta(\tau)\eta'(\tau)\right)
\Big]\ , \\[4pt]
& =\gamma
\frac{(2\pi)^2}{\beta}
\sum_{n\ge 0}n^2\left[ 
2T_0|\varepsilon_n|^2+
\Big(P_0+\frac{2\pi n}{\beta}\Big)\varepsilon_{n}\sigma_{n}^\ast
+
\Big(P_0-\frac{2\pi n}{\beta}\Big)\varepsilon_{n}^\ast\sigma_n
\right]+\\[4pt]
&\hspace{12pt} +\gamma(8\pi)\sum_{n> 0}n\left[\left(\frac{2\pi n}{\beta}\right)^2-\left(\frac{P_0}{2}\right)^2\right]\eta_n\eta_n^\ast\ ,
\end{aligned}
\end{equation}
where in the second equality we used the Fourier decompositions of the functions (\ref{eq:Fourier}). Since we are now interested in the disk, we should evaluate this at $(P_0,T_0)=\frac{2\pi}{\beta}(1,0)$. In this case, the degenerate directions of the symplectic form (\ref{eq:81}) for the disk are enhanced to $(\varepsilon_0,\sigma_0,\varepsilon_1^\ast,\sigma_1)$ and $(\eta_{1/2},\eta_{1/2}^\ast)$ in the bosonic and fermionic sectors respectively, so that the measure appearing in (\ref{eq:83}) is
\begin{equation}
d\varepsilon_1d\sigma^\ast_1(4\pi \gamma^2c_0 P_0)
\prod_{n\ge 2}
d^2\varepsilon_nd^2\sigma_n
(2\pi  \gamma^2c_0 P_0)^2
n^2(n^2-1)
\prod_{m\ge \frac{3}{2}} 
\frac{d^2\eta_m}{\beta \gamma^2 c_0 P_0^2(4m^2-1)}\ ,
\end{equation}
where $d^2\varepsilon_n=d\varepsilon_nd\varepsilon_n^\ast$. Putting everything together in (\ref{eq:83}) we can solve all the integrals and obtain the final expression for the disk partition function
\begin{equation}
Z_{\rm disk}(\beta)=
(4c_0\gamma \beta)\prod_{n\ge 2}\left(\frac{c_0\gamma \beta }{n}\right)^2
\prod_{m\ge \frac{3}{2}}
\left(
\frac{2\pi m}{c_0 \gamma \beta} \right)=
\frac{4(c_0 \gamma \beta)}{2\pi(c_0 \gamma \beta)^3}
\frac{\sqrt{2}(c_0\gamma \beta)}{\pi}=
\frac{2\sqrt{2}}{(c_0\pi^2) \gamma \beta}
\ ,
\end{equation}
where we have used the Riemann and Hurwitz Zeta functions to regularize the infinite product. Since the on-shell action of the disk vanishes, the only $\beta$ dependence comes from the one-loop determinant, which scales as predicted by the simple counting of the zero modes indicated in (\ref{eq:ZeroModes}). Conveniently setting $c_0=1/\pi^2$ and rescaling $\beta\rightarrow \beta/\gamma$ in order to have a dimensionless inverse temperature, we recover the result quoted in (\ref{eq:IntroPart}).

\subsubsection*{Cylinder Partition Function}

The cylinder metric of circumference $b\in \mathbb{R}_+$ is given by $ds^2=dz^2+b^2d\varphi^2$ with $z\in \mathbb{R}$ and $\varphi\sim \varphi+1$. It can be written in Bondi gauge by changing coordinates to $r=(b/\beta)z$ and $\tau=\beta(\varphi-iz/b)$, which results in the Euclidean version of the metric (\ref{eq:LorBondi}) with $(P_0,T_0)=\frac{b^2}{2\beta^2}(0,1)$.\footnote{Note the inverse temperature $\beta$ is not defined as the circumference of the boundary circle but instead as the period of the coordinate $\tau$, obtained from analytically continuing the Bondi time $u\rightarrow i\tau$. See \cite{Kar:2022sdc} for a detailed discussion on this feature, which is an important difference of these models of flat space holography when compared to the standard AdS/CFT correspondence.} While this configuration certainly satisfies the equations of motion for arbitrary $H_0$, it is not an actual solution of the gravitational theory given that it cannot satisfy the boundary conditions on both boundaries (see Section B.1 of \cite{Kar:2022sdc}). This means the calculation of the cylinder partition function is slightly more subtle than the case of the disk (\ref{eq:83}).

Similarly as in JT gravity \cite{Saad:2019lba}, one can bypass this problem by constructing the cylinder path integral by appropriately gluing two ``half-cylinders". In the region where the half-cylinders meet at $z=0$, the metric is locally given by
\begin{equation}\label{eq:84}
ds^2=dz^2+(bd\varphi+\upsilon \delta(z)dz)^2\ .
\end{equation}
There are two moduli $(b,\upsilon)$ associated to the gluing that one must integrate over: the circumference~$b$ and the relative twist $\upsilon$ between the two boundaries that are being glued. The correct measure over this two-dimensional moduli space can be determined from the symplectic form provided by the BF theory (\ref{eq:SympForm}). To do so, we must first construct the flat $\boldsymbol{F}=0$ gauge connection $\boldsymbol{A}$ associated to the metric (\ref{eq:84}). A straightforward computation shows $\boldsymbol{A}$ is given by
\begin{equation}
\boldsymbol{A}=
\frac{1}{\sqrt{2}}
\big[(bd\varphi+\upsilon \delta(z)dz)- i dz\big]
P_++
\frac{1}{\sqrt{2}}
\big[(bd\varphi+\upsilon \delta(z)dz)+ i dz\big]P_-+ibzd\varphi I+\psi^-Q_-\ ,
\end{equation}
where $\psi^-$ satisfies $d\psi^-=0$. From this expression one can evaluate the symplectic form (\ref{eq:SympForm}) and find
\begin{equation}
\Omega(\delta_1\boldsymbol{A},\delta_2\boldsymbol{A})=\gamma^2 c_0  \int_{\mathcal{M}}
\left[
\delta_1b \delta_2\upsilon
-
\delta_1\upsilon 
\delta_2b
\right]
\delta(z) d\varphi  \wedge  dz
\ ,
\end{equation}
where the Dirac delta localizes the integral to the gluing region $z=0$. In this way, one finds the symplectic form associated to the gluing of two half-cylinders is given by
\begin{equation}
\Omega=\gamma^2 c_0 \,\delta b \wedge \delta \upsilon\ ,
\end{equation}
which is the same as the one obtained for the JT and CJ gravity theories \cite{Saad:2019lba,Kar:2022sdc}. Putting everything together, the cylinder partition function can be computed from 
\begin{equation}\label{eq:85}
Z_{\rm cylinder}(\beta_1,\beta_2)=2\gamma^2 c_0\int_0^{\infty}db\,b\,
Z_{\rm half-cylinder}(\beta_1,b)
Z_{\rm half-cylinder}(\beta_2,b)\ ,
\end{equation}
where we have solved the integral over the twist using the half-cylinder partition functions are independent of $\upsilon$. The additional factor of two comes from the sum over inequivalent bulk spin structures, see Section~2.4.3 of \cite{Stanford:2019vob}.\footnote{
Note the final result for the topological expansion of $\mathcal{N}=1$ CJ supergravity (\ref{eq:IntroPart}) is insensitive to whether we define the theory which sums or takes the differences between even and odd spins structures \cite{Stanford:2019vob}. As a result, both bulk theories will be given described by the same random matrix model, e.g. see section 5 of \cite{Balasubramanian:2020jhl}.}

The half-cylinder partition function can be obtained from a one-loop computation as in (\ref{eq:83}), where the quadratic fluctuations are around the on-shell solution $(P_0,T_0,H_0)=\frac{b^2}{2\beta^2}(0,1,0)$. For these values, the symplectic form associated to the asymptotic degrees of freedom (\ref{eq:81}) contains only two bosonic degenerate directions $(\varepsilon_0,\sigma_0)$, so that the measure obtained from the Pfaffian is
\begin{equation}
\prod_{n\ge 1}
d^2\varepsilon_nd^2\sigma_n
\left[
\frac{c_0(2\pi n \gamma)^2}{\beta}
\right]^2
\prod_{m\ge \frac{1}{2}} 
d^2\eta_m
\left[ \frac{c_0(4\pi n\gamma)^2}{\beta}\right]^{-1}\ .
\end{equation}
Using that the on-shell action is $I_\partial^{\rm (on-shell)}=\gamma \beta T_0$ we can solve the relevant integrals of the quadratic action (\ref{eq:Quad}) and find
\begin{equation}
Z_{\rm half-cylinder}(\beta,b)=
e^{-\frac{\gamma b^2}{2\beta}}
\prod_{n\ge 1}\left(\frac{c_0\gamma \beta }{n}\right)^2
\prod_{m\ge \frac{1}{2}}
\left(
\frac{2\pi m}{c_0\gamma \beta}
\right)=
\frac{\sqrt{2}}{2\pi c_0\gamma \beta}
e^{-\frac{\gamma b^2}{2\beta}}\ ,
\end{equation}
where the infinite products are regularized similarly as for the disk. The scaling of this expression with $\beta$ is consistent with the general formula (\ref{eq:ZeroModes}), given that in this case there are only two bosonic zero modes. Using this in (\ref{eq:85}), we solve the integral over the gluing curve and arrive at the final expression for the cylinder partition function
\begin{equation}
Z_{\rm cylinder}(\beta_1,\beta_2)=
\frac{1}{c_0\pi^2}
\frac{1}{\gamma(\beta_1+\beta_2)}\ ,
\end{equation}
which gives (\ref{eq:IntroPart}) after fixing $c_0=1/\pi^2$ and rescaling $\beta_i$ by $1/\gamma$. Notice that although we are free to set the proportionality constant $c_0$ to any value, it cannot be absorbed into a redefinition of $S_0$ in (\ref{eq:TopExp2}), as is the case for JT gravity (see Section 3.4 in \cite{Saad:2019lba}). This is because, as in ordinary CJ gravity, $c_0$ does not scale with the genus and number of boundaries in the same way as $e^{-S_0}$.

\section{Dual Random Matrix Model}
\label{sec:3}

This section contains all the analysis and computations regarding the random matrix model dual to CJ supergravity. We show there is a unique double scaled model that reproduces the topological expansion (\ref{eq:TopExp2}), and then assume the non-perturbative completion provided by the matrix model to exactly compute several observables of the gravitational theory.

\subsection{Topological Expansion from Loop Equations}
\label{sec:3.1}

To determine the appropriate random matrix model there are three pieces of data one needs to fix: 
\vspace{8pt}
\begin{enumerate}
\setlength\itemsep{4pt}
   \item Symmetry class of the random matrix $M$. \label{item:1}
   \item Matrix operator $\mathbb{O}(\beta)$ corresponding to the insertion of an asymptotic boundary in gravity. \label{item:2}
   \item Probability measure over the ensemble. \label{item:3}
\end{enumerate}
\vspace{8pt}
In this subsection we explain how this data is uniquely fixed in order to match with the topological expansion of the Euclidean partition function (\ref{eq:IntroPart}), effectively deriving the identity in (\ref{eq:IntroMatching}).

Let us start by considering the first item. There exist ten standard symmetry classes of random matrix models. The three $\boldsymbol{\beta}=1,2,4$ Dyson ensembles \cite{Dyson:1962es} correspond to a matrix $M$ that is real symmetric, complex Hermitian, or quaternionic Hermitian, while the remaining seven are the $(\boldsymbol{\alpha},\boldsymbol{\beta})$ ensembles of Altland and Zirnbauer~\cite{1997}. Observables in all these models can be studied perturbatively in a large $N$ 't Hooft expansion. The loop equations \cite{Migdal:1984gj} is a method that allows one to compute such expansion to arbitrary order, via a set of recursion relations \cite{Eynard:2004mh,Stanford:2019vob}. This matrix model expansion is very much related, and ultimately identified \cite{DiFrancesco:1993cyw}, with the topological expansion of observables in two-dimensional quantum gravity~(\ref{eq:IntroTopological}). 

Out of the ten available ensembles, we can find the suitable one from some of the features exhibited by the CJ supergravity partition function. To start, only orientable surfaces are included in the Euclidean topological expansion. There are only two ensembles consistent with this \cite{Stanford:2019vob}: the $\boldsymbol{\beta}=2$ Dyson and $(\boldsymbol{\alpha},\boldsymbol{\beta})=(1,2)$ Altland-Zirnbauer  ensembles.\footnote{The Altland and Zirnbauer ensemble with $(\boldsymbol{\alpha},\boldsymbol{\beta})=(1,2)$ is also sometimes called a complex matrix model \cite{Morris:1990cq,Dalley:1991qg}, given that its probability measure can be constructed from an arbitrary complex matrix $M$ where the potential that determines the measure is only a function of the combination $MM^\dagger$.} In addition, the gravity path integral exhibits the cancellation of contributions from higher genus and multi-boundary surfaces (\ref{eq:DiskCylTopExp}). As shown in the general analysis of the loop equations in \cite{Stanford:2019vob}, this singles out the $\boldsymbol{\beta}=2$ Dyson ensemble, which can produce precisely such cancellations for a large class of models. This means the data in item \ref{item:1} must be fixed so that $M$ is a complex Hermitian matrix.

Consider then a random Hermitian squared matrix $M$ drawn from an ensemble whose probability density is determined by a potential $V(M)$. Observables $\mathcal{O}$ are arbitrary functions of $M$, such that their expectation values are computed as
\begin{equation}\label{eq:90}
\langle \mathcal{O} \rangle = 
\frac{1}{\mathcal{Z}}\int dM\,
\mathcal{O}\,e^{-N\,{\rm Tr}\,V(M)}
\ ,
\end{equation}
where $\mathcal{Z}$ is the numerator without the operator insertion and $dM$ the flat measure over the independent matrix components. An example of a very useful observable is the resolvent
\begin{equation}
W(z)={\rm Tr}\frac{1}{z-M}\ ,
\qquad \qquad
z\in \mathbb{C}\ ,
\end{equation}
which is a holomorphic function in the complex plane away from the spectrum of $M$. Its higher trace generalization is $W(z_1,\dots,z_n)=\prod_{i=1}^nW(z_i)$. Knowledge of this observable completely determines the expectation value of all other trace class operators. 

The connected expectation value of the resolvent can be written in a large $N$ series expansion
\begin{equation}\label{eq:87}
\langle W(z_1,\dots,z_n) \rangle_c \simeq \sum_{g=0}^{\infty}
N^{2(1-g)-n}W_g(z_1,\dots,z_n)\ ,
\end{equation}
where the symbol $\simeq$ indicates the right-hand side is missing non-perturbative contributions in $1/N$. The loop equations are a set of recursion relations that fully determine the expansion coefficients $W_g(z_1,\dots,z_n)$ given a particular potential $V(M)$ defining the matrix model probability density. In fact, instead of the potential, the loop equations are also completely determined by the leading large~$N$ behavior of the eigenvalue spectral density\footnote{Since it will be enough for our purposes, we are restricting to the single-cut case, in which the leading density is supported in a single interval.}
\begin{equation}\label{eq:88}
\lim_{N\rightarrow \infty} \frac{1}{N}
\langle {\rm Tr}\,\delta(\lambda-M) \rangle =
\frac{1}{2\pi}h(\lambda)\sqrt{(a_--\lambda)(\lambda-a_+)}
\times \boldsymbol{1}_{[a_-,a_+]}\ ,
\end{equation}
where $\boldsymbol{1}_A$ is the indicator function and $h(\lambda)$ an analytic function that is non-negative in $\lambda\in [a_-,a_+]$. From the data provided by $a_\pm$ and $h(\lambda)$, a standard computation allows one to determine the potential~$V(M)$ \cite{Eynard:2015aea}.

The nature of the loop equations changes substantially depending on whether the parameters~$a_\pm$ are finite or not. While for finite $a_\pm$ all the coefficients in the expansion (\ref{eq:87}) are generically non-zero, a dramatic cancellation of precisely the same kind as observed for the Euclidean partition function~(\ref{eq:DiskCylTopExp}), is recovered when $a_\pm\rightarrow\pm \infty$, i.e. only $W_0(z)$ and $W_0(z_1,z_2)$ are non-zero \cite{Stanford:2019vob}.\footnote{For a specific and more detailed discussion of how this mechanism arises from the matrix model loop equations, see Section 3.2 in \cite{Kar:2022sdc}, as well as Appendix C in that paper. As mentioned above, this kind of cancellation is not possible for the $(\boldsymbol{\alpha},\boldsymbol{\beta})=(1,2)$ Altland-Zirnbauer ensemble.} The correct and rigorous way to reach this regime is to instead take a double scaling limit, controlled by an additional parameter $\delta\rightarrow 0$ and taking $1/N=\hbar/ \delta^{\#}$, where the power of $\delta$ depends on the particular potential $V(M)$, while simultaneously rescaling the matrix to $\bar{M}=M/\delta$. This has the effect of zooming into the $\lambda\sim 0$ eigenvalues in (\ref{eq:88}). In this limit one should consider observables of the rescaled matrix $\bar{M}$, so that the large $N$ expansion of the resolvent in (\ref{eq:87}) is replaced by a small~$\hbar$ series (not related to Planck's constant). One finds the expansion for the resolvent of the matrix~$\bar{M}$ collapses to
\begin{equation}\label{eq:89}
\begin{aligned}
\langle W(z) \rangle & \simeq \frac{1}{\hbar} W_0(z)\ , \\[4pt]
\langle W(z_1,z_2) \rangle_c & \simeq W_0(z_1,z_2)\ , \\[4pt]
\langle W_0(z_1,\dots,z_n)\rangle_c & \simeq 0\ ,
\end{aligned}
\end{equation}
where $n\ge 3$. This expansion has precisely the required structure to match gravity~(\ref{eq:DiskCylTopExp}). One still needs to fix the data in items \ref{item:2} and \ref{item:3}, which is done as follows.

While $W_0(z)$ will depend on the fine grained details of the potential $V(M)$, this is not the case for $W_0(z_1,z_2)$, which is only determined by $a_\pm$. Since we have already fixed $a_\pm\rightarrow \pm \infty$, the function $W_0(z_1,z_2)$ can be unambiguously computed and written as \cite{Stanford:2019vob}
\begin{equation}
W_0(z_1,z_2)=
\begin{cases}
\qquad \,\, 0 \qquad 
\ , \qquad (z_1,z_2)\,\,{\rm same\,\,sheet}\ , \\
\displaystyle
\,\,\frac{-1}{(z_1-z_2)^2}
\ , \qquad (z_1,z_2) \,\,{\rm different\,\,sheets}\ .
\end{cases}
\end{equation}
This function is defined on a two-sheeted Riemann surface with a branch-cut along the entire real line. From this expression and (\ref{eq:89}) one can compute the full perturbative expansion of two insertions of an arbitrary trace class operator $\mathbb{O}(\beta)$. This allows us to fix item \ref{item:2}, by picking the operator which ensures the matching with the cylinder partition function $Z_{\rm cylinder}(\beta_1,\beta_2) \simeq \langle \mathbb{O}(\beta_1)\mathbb{O}(\beta_2) \rangle_c$. A straightforward computation (see Section 3.2 of \cite{Kar:2022sdc} for details) shows the required matrix operator is given by
\begin{equation}\label{eq:92}
\mathbb{O}(\beta)=\int_{-\infty}^{+\infty}dp\,{\rm Tr}\,e^{-\beta(\bar{M}^2+p^2)}\ .
\end{equation}
The integral over $p$ can be solved to give a factor of $\sqrt{\pi/\beta}$.

All we have left to do is fix the probability measure of the matrix model, as indicated in the item~\ref{item:3}. Instead of specifying a potential $V(M)$ in (\ref{eq:90}), it is more clear (and ultimately equivalent) to first determine the analytic function $h(\lambda)$ controlling the eigenvalue spectral density in (\ref{eq:88}). Defining $\alpha_i$ as the eigenvalues of the rescaled matrix $\bar{M}=M/\delta$ (in contrast to the ordinary eigenvalues $\lambda_i$ of $M$) the spectral density of $\bar{M}$ is
\begin{equation}
\rho(\alpha)={\rm Tr}\,\delta(\alpha-\bar{M})\ .
\end{equation}
Its leading behavior, which due to (\ref{eq:89}) it is actually the only non-zero perturbative contribution, is determined by $h(\lambda)$ in (\ref{eq:88}). In order to match with gravity $\langle \mathbb{O}(\beta) \rangle\simeq e^{S_0} Z_{\rm disk}(\beta)$ one needs to take
\begin{equation}\label{eq:93}
\langle \rho(\alpha) \rangle \simeq \frac{1}{\pi \hbar}\ ,
\qquad {\rm where} \qquad
\hbar = \frac{e^{-S_0}}{2\sqrt{2}}\ ,
\end{equation}
corresponding to $h(\lambda)$ constant. The potential $V(M)$ implied by this constant spectral density can be worked out and found to be the simplest case possible, a quadratic polynomial $V(M)=\frac{1}{2}M^2$. While in the naive large $N$ limit this gives Wigner's semi-circle law (\ref{eq:Wigner}), the appropriate double scaling limit is attained by zooming into the $\lambda\sim 0$ region in the following way
\begin{equation}\label{eq:double}
\frac{1}{N}=\hbar \delta \ ,
\qquad \qquad
\lambda_i=\alpha_i\delta\ ,
\end{equation}
with $\delta\rightarrow 0$. All in all, we have shown there is a unique double scaled matrix model which is able to reproduce the CJ supergravity partition function to all orders in perturbation theory
\begin{equation}\label{eq:91}
Z(\beta_1,\dots,\beta_n)\simeq \langle \mathbb{O}(\beta_1)\dots\mathbb{O}(\beta_n) \rangle_c\ .
\end{equation}
Let us now make a number of comments about certain features and subtleties of the matching between gravity and matrix model that we have just derived.

\paragraph{Uniqueness and Double Scaling:} As showed above, there is a single double scaled matrix model which ensures the matching in (\ref{eq:91}). When making this statement, the words ``double scaled" matrix model are important, as there is actually always an inherent ambiguity in the double scaling limit. This comes from the fact it involves zooming into eigenvalues with $\lambda\sim 0$, essentially forgetting about the behavior of large magnitude eigenvalues. More concretely, there is always an infinite class of matrix model potentials $V(M)$ one could have chosen which result in the same double scaled model. Given that the particular representative potential one picks to perform calculations is inconsequential, one usually takes the simplest one, in this case $V(M)=\frac{1}{2}M^2$. This is very much related to the universality of eigenvalue repulsion in matrix models.

\paragraph{Was it bound to work?:} From the way we have constructed the matrix model it might seem that once the gravitational partition function takes the form in (\ref{eq:DiskCylTopExp}), one is always going to find an appropriate random matrix model that does the job. One could think that no matter the specific details of the disk and cylinder partition functions, one can always pick the operator $\mathbb{O}(\beta)$ and the leading spectral density in order to ensure the matching. This is actually not the case. Although one can always find an operator $\mathbb{O}(\beta)$ such that the cylinder partition function is reproduced, in certain cases the matching with the disk topology might require a spectral density that is simply not allowed by the matrix model. By this we mean one might find the function $h(\lambda)$ in (\ref{eq:88}) that is required for the matching is not analytic or non-negative. As an example, if the cylinder partition function of CJ supergravity stays the same while the disk scales with the inverse temperature with a half-integer power, no matrix model can reproduce the corresponding partition functions. There is an underlying relation between the disk and cylinder partition functions of CJ supergravity that enables the matching to work.

\paragraph{Free Particle Sector:} Given what we have learned from Euclidean partition functions in AdS quantum gravity, one could have guessed the operator $\mathbb{O}(\beta)$ would take the form ``~${\rm Tr}\,e^{-\beta H}$~". Here, the operator $H$ should be interpreted as the generator of translations along the direction that was analytically continued to define the path integral. Since in this case it is the Bondi time $u\rightarrow i\tau$ in (\ref{eq:LorBondi}), we should think of $H$ as the Bondi Hamiltonian. The operator $\mathbb{O}(\beta)$ in (\ref{eq:92}) has precisely the right structure, with the surprising feature that the Bondi Hamiltonian contains not only a discrete contribution coming from the eigenvalues of the matrix $\bar{M}^2$, but also a continuous sector $p^2$ corresponding to the energy of a free particle. In fact, precisely the same operator $H$ was obtained for ordinary CJ gravity \cite{Kar:2022sdc,Kar:2022vqy}, showing it is not an accident but a persistent feature of these theories of flat quantum gravity. What is the origin and significance of this peculiar structure? While we do not have a definite answer for this question, there are a few comments we can make. 

At the technical level, one can trace the origin of the free particle to the central extension $I$ that is present in the Maxwell superalgebra (\ref{eq:MaxwellSuperalgebra}). The $\beta$ scaling of the partition function (\ref{eq:ZeroModes}) is directly influenced by this central extension, since it results in an additional zero mode which provides an extra factor of $1/\sqrt{\beta}$ to the disk and cylinder partition functions. This is precisely the contribution of the free particle to the matrix operator $\mathbb{O}(\beta)$ in (\ref{eq:92}). 

This suggests one might be able to remove the free particle by getting rid of the central extension. By removing it, the CJ gravity theory is replaced by a BF theory whose gauge group is ordinary Poincaré, which is nothing more than ordinary flat JT gravity. This is not good, as flat JT gravity only has a thermal solution with a single fixed (infinite) temperature, and is therefore not amenable to the thermodynamic analysis implied by the computation of $Z(\beta)$.\footnote{A more detailed discussion of this issue can be found in \cite{Stanford:2020qhm,Godet:2021cdl}, which also considers the very much related case of the CGHS gravity theory \cite{Callan:1992rs}. Although the analysis of \cite{Afshar:2021qvi} is able to bypass this problem by picking a different set of boundary conditions, their formulation of the theory does not seem to allow for a cylinder topology, which is crucial for the matrix model interpretation.} As a result, the central extension of the algebra seems to be very much required, playing a crucial role in the formulation of the theory.

Another perspective is that the free particle sector is not a bug but a feature of flat space quantum gravity. A feasible interpretation might be that the free particle, which yields a continum spectrum for the Bondi Hamiltonian, is somehow related to the infinite volume of flat space. This is in contrast to the covariant AdS ``box", which has a discrete spectrum instead. Although conceptually appealing, further evidence needs to be gathered to make such statement concrete.

\subsection{Non-perturbative Completion}
\label{sec:3.2}

In our discussion so far the analysis of both the matrix model and supergravity has been limited to perturbative effects in the parameter $\hbar\propto e^{-S_0}$. What about non-perturbative contributions? While on the gravitational side there is currently no known method for capturing such effects, a non-perturbative analysis is indeed tractable for the matrix model. In this subsection we assume the non-perturbative completion of the gravity theory provided by the matrix model and use it to investigate CJ supergravity non-perturbatively. More concretely, we assume we can replace the symbol $\simeq$ in (\ref{eq:91}) by a strict equality. The method of loop equations is an intrisic perturbative approach and therefore not useful for computing non-perturbative contributions in the matrix model. Instead, we shall use the method of orthogonal polynomials, that is better suited for this task.\footnote{See Section 5 of \cite{Eynard:2015aea} and Appendix C in \cite{Rosso:2021orf} for a more detailed discussion of this method for finite $N$ and double scaled models, respectively.} 

The central quantity that determines all observables is the matrix model kernel $K(\lambda,\lambda')$. For a Hermitian matrix model with an arbitrary potential $V(M)$, the computation of the kernel is quite complicated and can be rarely performed analytically. However, for the Gaussian matrix model $V(M)=\frac{1}{2}M^2$ one can write the kernel explicitly for finite $N$. Moreover, in the double scaling limit, given by (\ref{eq:double}), it simplifies to the well know sine kernel
\begin{equation}\label{eq:Kernel}
K(\alpha,\alpha')=\frac{1}{\pi}
\frac{\sin\left[(\alpha-\alpha')/\hbar\right]}{\alpha-\alpha'}\ .
\end{equation}
In Appendix \ref{zapp:2} we explain in detail how to derive this result using the method of orthogonal polynomials. We should stress that in this context, the sine kernel is not an approximation. Instead, it is the exact result, including all perturbative and non-perturbative contributions in $\hbar$, to the kernel of the double scaled matrix model required to describe $\mathcal{N}=1$ CJ supergravity. As a result, all observables obtained from this kernel are exact.

\subsubsection{Bondi Spectrum}

One of the simplest observables is the spectral density $\varrho(E)$ of the supergravity theory, obtained from the inverse Laplace transform of the single boundary partition function. Since the path integral is defined through the analytic continuation of the Bondi time (\ref{eq:LorBondi}), $\varrho(E)$ captures the spectrum of the Bondi Hamiltonian. To all orders in perturbation theory, this is easily obtained from (\ref{eq:IntroPart}), which gives
\begin{equation}\label{eq:BondiSpec}
   \varrho(E)\simeq e^{S_0}2\sqrt{2}\Theta(E)\ .
\end{equation}
Non-perturbative contributions not captured by this expression can be obtained from the eigenvalue spectral density of the matrix model $\rho(\alpha)$, whose expectation value is determined by the diagonal components of the kernel
(\ref{eq:Kernel}). While for a generic matrix model one expects small non-perturbative oscillations around the leading perturbative behavior, this is not the case for the double scaled Gaussian model, where one finds no corrections of any kind ${\langle \rho(\alpha) \rangle=\frac{1}{\pi \hbar}}$. This means the non-perturbative completion provided by the matrix model predicts the constant Bondi spectrum in (\ref{eq:BondiSpec}) is actually exact, i.e. there is a strict equality.

\begin{figure}
    \centering
    \includegraphics[scale=0.60]{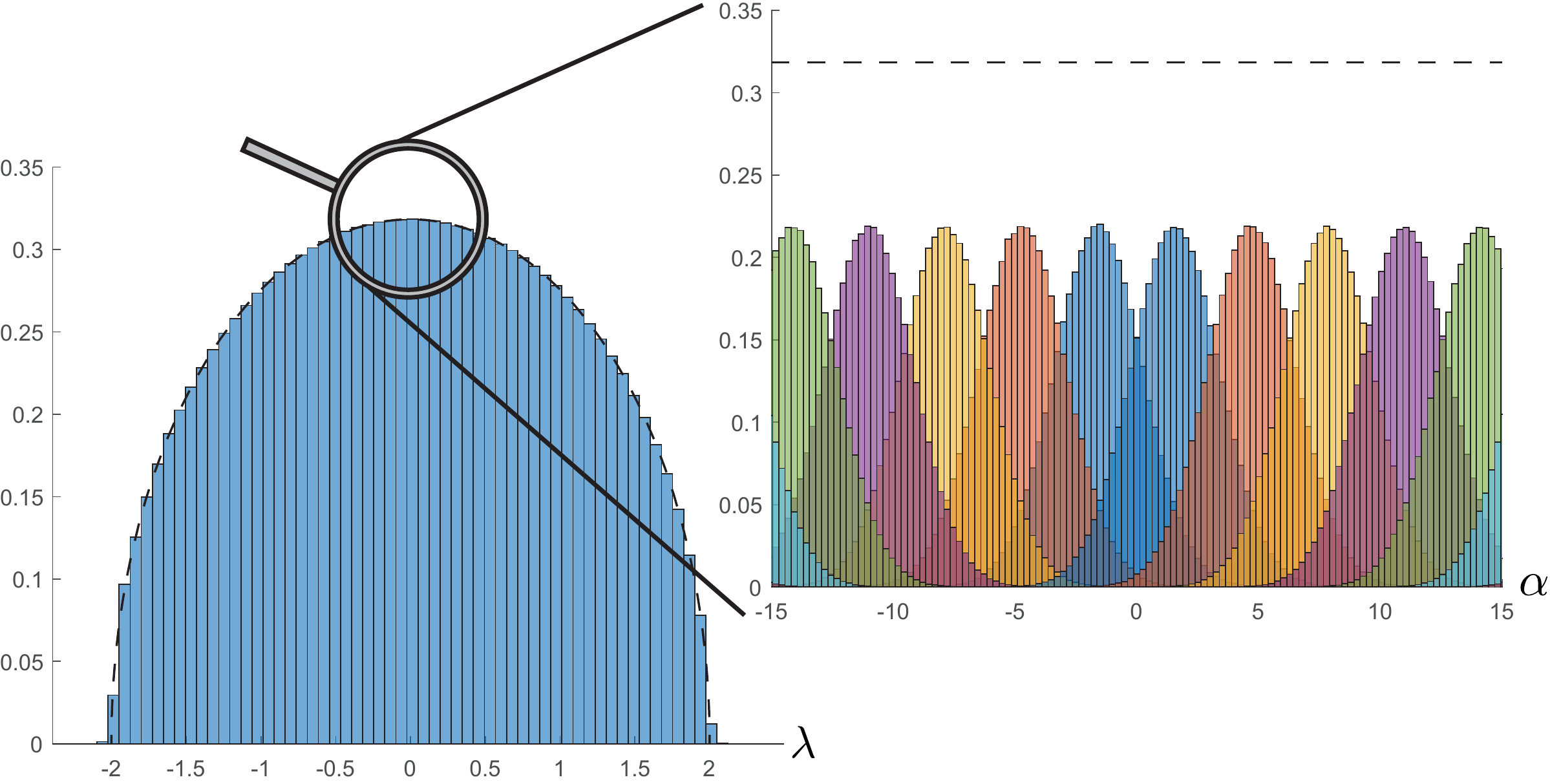}
    \caption{In the left diagram observe the histogram corresponding to the eigenvalues of $10^6$ Hermitian random matrices of size $N=100$, drawn from the Gaussian ensemble. The dashed line corresponds to Wigner's semi circle law (\ref{eq:Wigner}). To the right, we zoom-in the $\lambda\sim 0$ region by taking the double scaling limit $\alpha_i=\lambda_i \hbar N$ (\ref{eq:double}) with $\hbar=1$. The histograms in this second plot correspond to the ordered eigenvalues of each sample matrix.}
    \label{fig:1}
\end{figure}

A more fine grained characterization of the Bondi spectrum is obtained by studying the probability density function $p_i(\alpha)$ of individual eigenvalues. For an arbitrary matrix model, this is computed from a Fredholm determinant of an integral operator constructed from the matrix model kernel, which can be evaluated numerically (see \cite{Johnson:2021zuo,Bornemann_2009,Johnson:2022wsr,Kar:2022sdc}). In this case however, there is an alternative path that is simpler, as one can instead numerically sample an ensemble of Gaussianly distributed Hermitian random matrices and directly compute the average of any quantity of interest. More explicitly, one writes $M=(A+A^\dagger)/2\sqrt{N}$ with $A\in \mathbb{C}^{N\times N}$ a random matrix drawn from a normal distribution of unit variance and zero mean. After generating $10^6$ samples of size $N=100$, one computes their respective eigenvalues and obtains the histogram appearing in the left plot of Figure~\ref{fig:1}, which is nothing more than Wigner's semi-circle law (dashed line). The double scaling limit is obtained by zooming-in the region $\lambda\sim 0$ via the rescaling $\lambda_i=\alpha_i/\hbar N$ of the matrix eigenvalues. In the right diagram of the same figure, we plot the histograms of the individual ordered and rescaled eigenvalues of the sample matrices, which provide a good numerical approximation of the probability density functions $p_i(\alpha)$ of each eigenvalue. As expected, summing those histograms gives the correct constant value for the eigenvalue spectral density $\langle \rho(\alpha) \rangle=\frac{1}{\pi \hbar}$, without any non-perturbative oscillations.\footnote{If instead of zooming into the $\lambda\sim 0$ eigenvalues one takes a double scaling limit which focuses on the edge of Wigner's semi-circle, one obtains the Airy model. In that case, the eigenvalue spectral density does exhibit non-perturbative oscillations around the leading perturbative result, see Figure 3 of \cite{Johnson:2021rsh}.}

\begin{figure}
    \centering
    \includegraphics[scale=0.60]{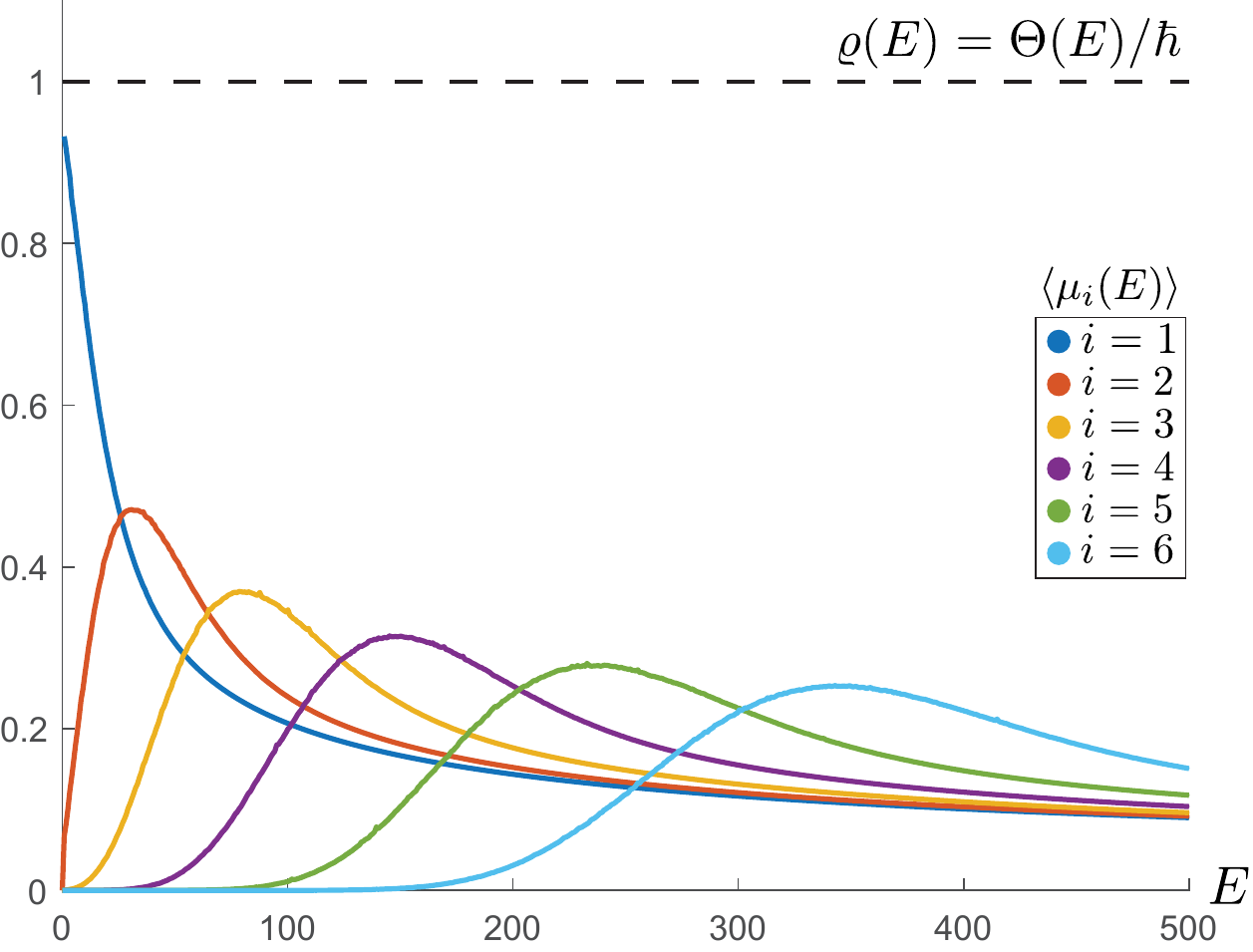}
    \caption{Expectation value of $\mu_i(E)$, defined in (\ref{eq:100}), for the first few values of $i$, computed from the sample of random matrices used to generate the plots in Figure \ref{fig:1}. Taking the sum of these curves one recovers the spectral density $\varrho(E)=\Theta(E)/\hbar$ with $\hbar=1$.}
    \label{fig:2}
\end{figure}

We can use this results to characterize the fine grained structure of $\varrho(E)$. Using the matching with the matrix model provided by $\mathbb{O}(\beta)$, one can derive the following formula for the Bondi spectral density
\begin{equation}\label{eq:100}
\varrho(E)=
\sum_{i=1}^{\infty}
\Big\langle 
\frac{\Theta(E-\alpha_i^2)}{\sqrt{\smash[b]{E-\alpha_i^2}}}
\Big\rangle \equiv \sum_{i=1}^{\infty}\langle \mu_i(E) \rangle \ ,
\end{equation}
where we have defined $\mu_i(E)$. Since $1/\sqrt{E}$ is the spectral density of a free particle, $\varrho(E)$ is constructed from a superposition of free particle spectral densities centered at the location of the eigenvalues $\alpha_i$ of the random matrix. While the whole sum can be computed analytically, giving the constant value in (\ref{eq:BondiSpec}), we can do better and use the sampling over the Gaussian random matrices to calculate each of the terms individually. In Figure \ref{fig:2} we plot $\langle\mu_i(E) \rangle$ for the first few values of $i$. Summing these contributions one recovers (as expected) the correct spectral density $\varrho(E)=\Theta(E)/\hbar$. A very similar structure for the fine grained Bondi spectrum was obtained for ordinary CJ gravity in \cite{Kar:2022vqy,Kar:2022sdc}, where the analogous curves in Figure \ref{fig:2} where computed using the Fredholm determinant approach instead.

\subsubsection{Multi-boundary observables}

While single boundary observables do not receive any kind of non-perturbative corrections, this is not the case for multi-boundary observables. As an example, consider the spectral form factor, defined from the Euclidean partition function with two boundaries as \cite{Cotler:2016fpe}
\begin{equation}\label{eq:102}
S(\beta,t)=Z(\beta+it)Z(\beta-it)+Z(\beta+it,\beta-it)\ .
\end{equation}
The first and second terms get contributions from disconnected and connected geometries respectively. Using (\ref{eq:IntroPart}) one can easily compute $S(\beta,t)$ to all orders in the small $\hbar$ perturbative expansion and find
\begin{equation}\label{eq:101}
S(\beta,t)\simeq \frac{1}{\hbar^2}
\frac{1}{\beta^2+t^2}
+
\frac{1}{2\beta}\ .
\end{equation}
Non-perturbative corrections to this expression can be obtained using the matrix model completion. This requires knowledge of the observable $\langle \mathbb{O}(\beta+it)\mathbb{O}(\beta-it) \rangle$, which can be calculated from 
\begin{equation}
\langle \rho(\alpha)\rho(\alpha') \rangle_c=\langle \rho(\alpha) \rangle\delta(\alpha-\alpha')-
K(\alpha,\alpha')^2\ ,
\end{equation}
where the kernel is given in (\ref{eq:Kernel}). Using this, one obtains the following exact expression for the spectral form factor
\begin{equation}\label{eq:103}
S(\beta,t)=
\frac{1}{\hbar^2}
\frac{1}{\beta^2+t^2}+
\frac{1}{2\beta}
\Big(1-e^{-\frac{1}{\hbar^2}\frac{2\beta}{\beta^2+t^2}}\Big)
+
\frac{1}{\hbar}
\sqrt{\frac{\pi}{2\beta(\beta^2+t^2)}}
\bigg(
1-{\rm Erf}\bigg[\frac{1}{\hbar}
\sqrt{\frac{2\beta}{\beta^2+t^2}}
\bigg]
\bigg)
\ ,
\end{equation}
which corrects the perturbative answer (\ref{eq:101}) in an interesting and non-trivial way.\footnote{Although the small $\hbar$ expansion only captures very rough features of the spectral form factor, note the large $\hbar$ expansion actually has an infinite radius of convergence.} It is quite remarkable that one can analytically write down the exact expression for the spectral form factor of this gravitational theory.

\begin{figure}
    \centering
    \includegraphics[scale=0.48]{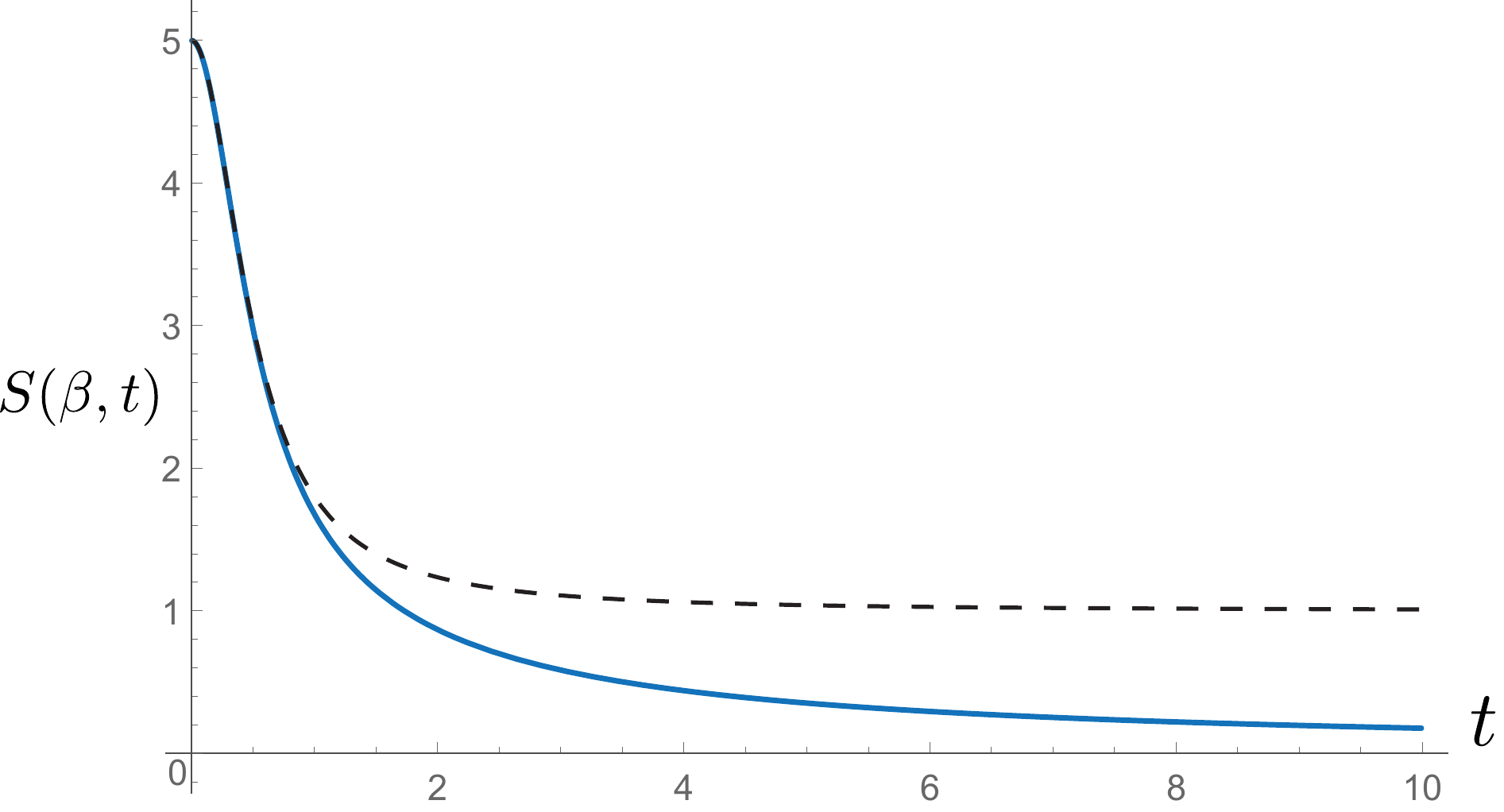}
    \caption{Spectral form factor (\ref{eq:102}) of $\mathcal{N}=1$ CJ supergravity at fixed $\beta=1/2$. The dashed and solid curves correspond to the perturbative (\ref{eq:101}) and exact (\ref{eq:103}) results respectively. The exact result decays to zero at late times, as expected from the continuous spectrum of $\mathbb{O}(\beta)$ in (\ref{eq:92}).}
    \label{fig:3}
\end{figure}

Let us analyze the time dependence of the spectral form factor for fixed $\beta$, starting with the perturbative expression in (\ref{eq:101}). At early times $t\ll e^{S_0}$, the contribution from the two disconnected boundaries dominates, as it is enhanced by a factor of $1/\hbar^2\propto e^{2S_0}$. This results in the initial ``dip" that is generally expected for the spectral form factor \cite{Cotler:2016fpe}. Later, for $t\sim e^{S_0}$, both terms are of the same order and the perturbative result for $S(\beta,t)$ in (\ref{eq:101}) takes a constant value. This is a departure from the more familiar case of AdS gravity, where for this time scale one instead finds a linear ``ramp" \cite{Saad:2018bqo}, related to the repulsion of the underlying microscopic spectrum of the theory. Given the non-perturbative completion of CJ supergravity found here, it should not be a surprise the ramp has disappeared, given that the repulsion of the eigenvalues of the matrix $\bar{M}$ is contaminated by the continuous free particle contribution to the spectrum of $\mathbb{O}(\beta)$ in (\ref{eq:IntroOperator}). For late times $t\gg e^{S_0}$, the perturbative expansion in (\ref{eq:101}) breaks down and one needs to consider the exact expression in (\ref{eq:103}) instead, which decays to zero.\footnote{See \cite{Blommaert:2022lbh} for a recent attempt of capturing the plateau from a perturbative analysis of the gravitational path integral.} While this behavior is again different when compared to the AdS case, where one instead gets a constant ``plateau" \cite{Cotler:2016fpe}, the late time decay observed here is exactly the expected behavior for a model with a continuous spectrum. In Figure \ref{fig:3} we plot the spectral form factor, the dashed and solid curves corresponding to the perturbative (\ref{eq:101}) and exact (\ref{eq:103}) results respectively. The overall behavior of the spectral form factor is analogous to the one observed for ordinary CJ gravity in \cite{Kar:2022sdc}.

A similar analysis can be performed for any other multi-boundary observable. One could argue the situation is even more interesting for three or more boundaries, since in those cases non-perturbative effects are the only non-zero contributions to the partition function (\ref{eq:IntroPart}). To illustrate this point, consider the partition function with three boundaries, so that non-perturbative effects as captured by the matrix model can be obtained from 
\begin{equation}
\begin{aligned}
\langle \prod_{i=1}^3 \rho(\alpha_i)\rangle_c & =
\langle \rho(\alpha_1)\rho(\alpha_2) \rangle_c\, \delta(\alpha_2-\alpha_3)
+
\langle \rho(\alpha_1)\rho(\alpha_3) \rangle_c\,
\delta(\alpha_1-\alpha_2)
+
\langle \rho(\alpha_2)\rho(\alpha_3) \rangle_c\,
\delta(\alpha_3-\alpha_1)+\\
& \hspace{10pt}
-2\langle \rho(\alpha_1) \rangle \delta(\alpha_1-\alpha_2)\delta(\alpha_1-\alpha_3)
+2K(\alpha_1,\alpha_2)K(\alpha_2,\alpha_3)K(\alpha_3,\alpha_1)
\ .
\end{aligned}
\end{equation}
This expression can be derived using the procedure explained in Appendix C of \cite{Rosso:2021orf}. From this one computes the connected expectation value of three insertions of $\mathbb{O}(\beta_i)$ and obtains the following expression for the three boundary partition function
\begin{equation}\label{eq:105}
Z(\beta,\beta,\beta)=
-\frac{2}{3}\frac{1}{\hbar\beta}
+
\frac{1}{\beta}
\left(1-e^{-\frac{3}{2\hbar^2\beta}}\right)
+
\frac{1}{\hbar \beta}\sqrt{\frac{3\pi}{2\beta}}
\left(
1-{\rm Erf}\left[\frac{1}{\hbar}\sqrt{\frac{3}{2\beta}}\right]
\right)
+I(\beta)\ ,
\end{equation}
where for convenience we have fixed all three boundaries to the same value of $\beta$ and defined
\begin{equation}
I(\beta)=
\frac{2}{(\pi\beta)^{3/2}}
\int_{-\infty}^{+\infty}d\alpha_1d\alpha_2d\alpha_3
\frac{\sin(\frac{\alpha_1-\alpha_2}{\hbar})}{\alpha_1-\alpha_2}
\frac{\sin(\frac{\alpha_2-\alpha_3}{\hbar})}{\alpha_2-\alpha_3}
\frac{\sin(\frac{\alpha_3-\alpha_1}{\hbar})}{\alpha_3-\alpha_1}
e^{-\beta(\alpha_1^2+ \alpha_2^2+\alpha_3^2)}\ .
\end{equation}
Solving this triple integral is quite challenging and we have not been able to find an analytic solution. It is therefore not possible to extract the perturbative contribution for small $\hbar$, given that one should first solve the integral and only then peform the expansion. However, there is no obstruction to first expanding for large $\hbar$ and then solving the integral. In this case, one finds the first few terms for the regime in which $\hbar\gg 1$ are given by
\begin{equation}
Z(\beta,\beta,\beta)=
\frac{1}{\hbar \beta}
\left[
\sqrt{\frac{3\pi}{2\beta}}-
\frac{2}{3}
\right]
-\frac{3}{2(\hbar\beta)^2}
+\frac{2}{(\hbar\beta)^3}
+\mathcal{O}(1/\hbar^4)
\ .
\end{equation}
This shows there are non-zero contributions to the multi-boundary partition functions that are not captured by the naive topological expansion in gravity.

\subsubsection{Low Temperature Thermodynamics}

Let us now turn our attention to the thermodynamics of CJ supergravity. The central quantity is the free energy, which in terms of the matrix model completion, is defined as
\begin{equation}\label{eq:104}
F_Q(T)=-T\langle \ln \mathbb{O}(1/T) \rangle\ .
\end{equation}
Calculating this observable is much more challenging than any of the previous, as it involves computing the ensemble average of the logarithm of the single trace operator $\mathbb{O}(\beta)$. A simple trick which simplifies the computation is to exchange these operations, and instead consider the ``annealed" free energy
\begin{equation}
F_A(T)=-T\ln \langle \mathbb{O}(1/T) \rangle\ ,
\end{equation}
as compared to the ``quenched" free energy in (\ref{eq:104}). Including all non-perturbative corrections, $F_A(T)$ can be computed exactly as
\begin{equation}\label{eq:Annealed}
F_A(T)=-T\ln T/\hbar\ .
\end{equation}
As first discussed in \cite{Engelhardt:2020qpv}, one should be careful with the annealed free energy, as it is a good approximation to the physically meaningful quenched free energy (\ref{eq:104}) only at high temperatures. Exchanging the ensemble average with the logarithm is not allowed at low temperatures. Therefore, one needs $F_Q(T)$ to fully characterize the thermodynamics of the gravitational theory.

\begin{figure}
    \centering
    \includegraphics[scale=0.65]{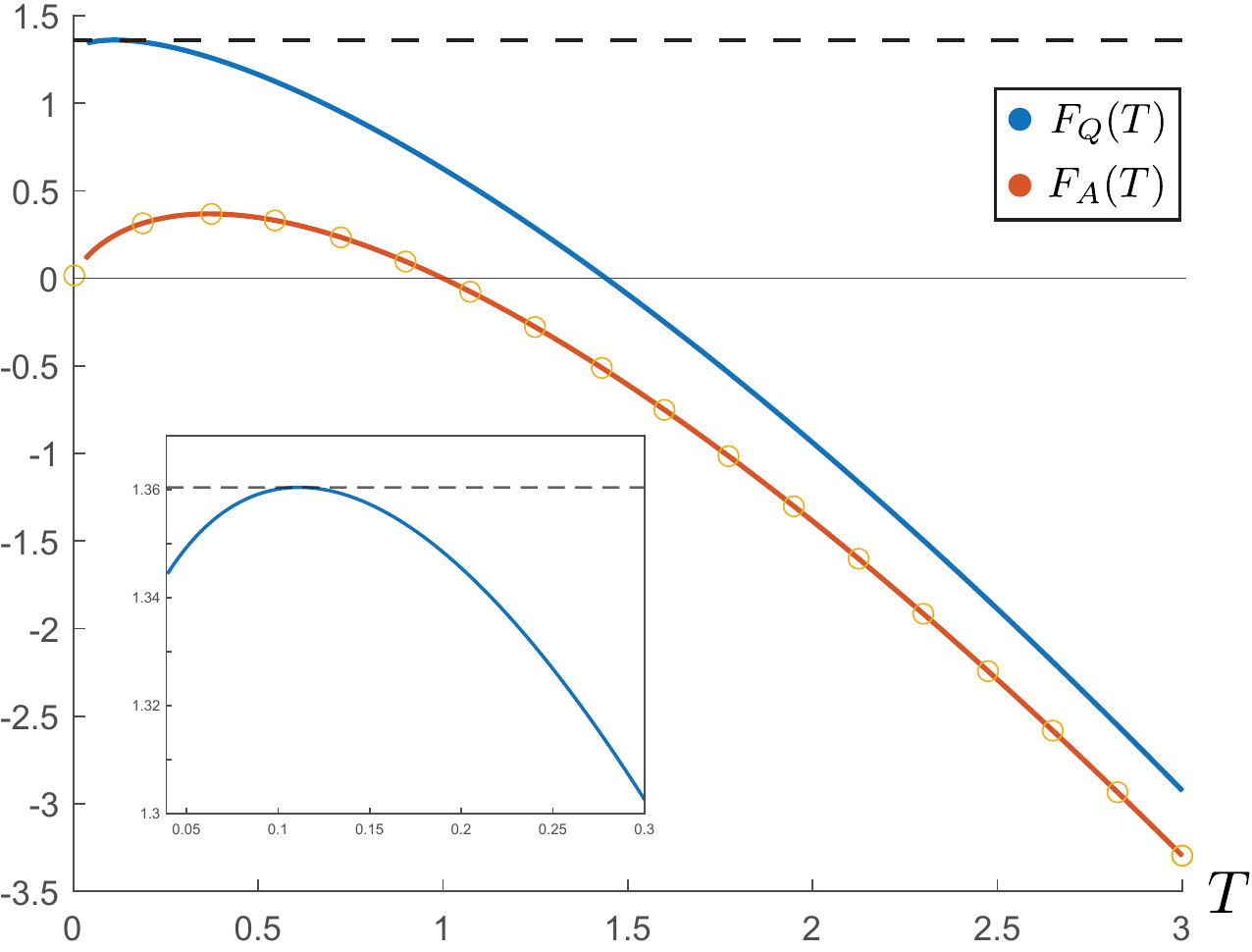}
    \caption{Quenched and annealed free energies of $\mathcal{N}=1$ CJ supergravity computed numerically from the sample of random matrices shown in Figure \ref{fig:1}, using the fourteen smallest magnitude eigenvalues of each matrix. The circled data shown in the plot corresponds to the analytic annealed free energy (\ref{eq:Annealed}), agreeing perfectly with the numerical result. The inset shows the ultra-low temperature behavior of the quenched free energy.}
    \label{fig:4}
\end{figure}

The simplest way of computing the quenched free energy is numerically, using the sample of random matrices shown in Figure \ref{fig:1}. At low temperatures, a good approximation is obtained by only considering the first few small magnitude eigenvalues of each matrix.\footnote{For the numerical precision required here, it is enough to include the first fourteen smallest magnitude eigenvalues of each matrix. Including more does not modify any of the features shown in Figure \ref{fig:4}. This approach for computing the quenched free energy was first proposed and implemented in \cite{Johnson:2021zuo,Johnson:2021rsh}.} In Figure \ref{fig:1} we plot both the quenched and annealed free energies computed in this way. As expected, we observe their difference becomes significant at low temperatures. The circled data corresponds to the analytic expression for the annealed free energy in (\ref{eq:Annealed}), showing perfect agreement with the numerical result. 

The thermodynamic entropy can be easily obtained from the free energy by simply taking a derivative $S(T)=-F'(T)$. Using the annealed free energy one finds $S(T)=1+\ln T/\hbar $, which is negative for $T<\hbar/e$ given that $F_A(T)$ is an increasing function at low temperatures (Figure \ref{fig:4}). Although in this regime one should use the quenched free energy instead, one finds the associated entropy still eventually becomes negative, since $F_Q(T)$ has a maximum (inset in Figure \ref{fig:4}). To better understand the origin of this behavior, it is convenient to separate the two contributions to $F_Q(T)$ coming from the discrete $\bar{M}^2$ and continuous $p^2$ parts of the spectrum of $\mathbb{O}(\beta)$
\begin{equation}
F_Q(T)=-T\langle \ln {\rm Tr}\,e^{-\bar{M}^2/T} \rangle
-\frac{1}{2} T\ln \pi T\ .
\end{equation}
The local maximum of the quenched free energy seen in Figure \ref{fig:4} is ultimately generated by the second term, i.e. by the continuous spectrum $p^2$. In fact, as analyzed in \cite{Kar:2022sdc}, any thermodynamic system with a continuous low temperature spectra will eventually result in a negative entropy. Given that the non-perturbative completion of CJ supergravity contains such a sector, it should not come as a surprise that the quenched free energy in Figure \ref{fig:4} has a maximum at ultra-low temperatures. The ultimate meaning and significance of this feature, which is also present for ordinary CJ gravity \cite{Kar:2022sdc}, is currently unclear.

\section{Final Remarks}
\label{sec:4}

In this work we have proposed an explicit realization of flat space holography in two space-time dimensions. The most remarkable feature of our construction is the simplicity of the boundary theory, given by an exactly solvable double scaled Gaussian matrix model. This is surprising, as in the bulk definition of $\mathcal{N}=1$ CJ supergravity (\ref{eq:SCJAction}) there is no obvious feature which hints towards the simplicity of the underlying holographic description. Ultimately, this theory seems like an ideal model where concrete questions that probe the nature of flat space quantum gravity may be answered.

In this regard, studying the S-matrix might be an interesting avenue to explore. Not only it is arguably the most natural observable in flat space gravity, but it might also allow us to make direct contact with the celestial holography program. However, the computation of scattering amplitudes requires developing additional technology, as the first step involves performing a careful analysis of the classical phase space of the theory, as performed in \cite{Kar:2022sdc} for ordinary CJ gravity. Furthermore, one also needs to understand how the theory looks when coupled to probe matter, as it is this additional matter sector that allows the creation of non-trivial asymptotic states that can ultimately scatter against each other. 

Without a doubt, the most puzzling aspect of our analysis is the continuous free particle spectrum present in the matrix operator $\mathbb{O}(\beta)$. In fact, this project was started from the desire to better understand such a sector, after it was previously encountered in ordinary CJ gravity \cite{Kar:2022vqy,Kar:2022sdc}. The free particle does not seem to be an accident but instead a robust feature of this class of two-dimensional flat space theories, unchanged by the addition of minimal supersymmetry. Exploring whether other extensions of CJ gravity which include unorientable surfaces or extended supersymmetry modify the operator $\mathbb{O}(\beta)$ in any way, is the natural next step in trying to better understand the degrees of freedom described by these theories. Moreover, although technically more challenging, studying the finite cut-off theory might help determine whether there is a concrete relation between the infinite volume of flat space and the continuous spectrum. We hope to revisit these questions in future work.

\noindent \paragraph{Acknowledgements:} It is a pleasure to thank Panos Betzios, Nikolay Bobev, Oscar Fuentealba, Hernán González, Arjun Kar, Lampros Lamprou, Charles Marteau, and Krzysztof Pilch for helpful comments and/or discussions. This work is supported in part by the National Sciences and Engineering
Research Council of Canada, and in part by the Simons Foundation.

\appendix

\section{Minimal Maxwell Superalgebra}
\label{zapp:1}

In this Appendix we further expand on certain aspects regarding the $\mathcal{N}=1$ supersymmetric extensions of the Maxwell algebra (\ref{eq:MaxwellAlgebra}). Several versions of the Maxwell superalgebra have been constructed \cite{Soroka:2004fj,Bonanos:2009wy,Bonanos:2010fw,Concha:2015woa}, mostly in dimensions higher than two. The superalgebra used in this work, which has not appeared previously in the literature, can be obtained from a In\"{o}n\"{u}-Wigner contraction of the $\mathfrak{osp}(1|2)$ superalgebra, which contains three bosonic $(J,P_\pm)$ and two fermionic $Q_\pm$ generators. The non-vanishing (anti-)commutators of $\mathfrak{osp}(1|2)$ are given by \cite{Frappat:1996pb}
\begin{equation}\label{eq:osp12}
\begin{aligned}
{\rm Bosonic:}& \qquad
\hspace{15pt} [J,P_\pm]=\pm P_\pm\ ,
\qquad
\hspace{30pt} [P_+,P_-]=2J\ , \\[4pt]
{\rm Mixed:}& \qquad
\hspace{15pt} [J,Q_\pm]=\pm\frac{1}{2}Q_\pm\ ,
\qquad  \hspace{17pt}
[P_\pm,Q_\mp]=-Q_\pm\ , \\[4pt]
{\rm Fermionic:}& \qquad
\lbrace Q_\pm,Q_\pm \rbrace =\pm \frac{1}{2} P_\pm\ ,
\qquad
\hspace{12pt} \lbrace Q_+,Q_- \rbrace=\frac{1}{2}J\ ,
\end{aligned}
\end{equation}
with the quadratic Casimir given by $C_2=\frac{1}{2}\lbrace P_+,P_- \rbrace+J^2-[Q_+,Q_-]$.

To start, the ordinary $\mathcal{N}=1$ Poincaré superalgebra can be obtained by rescaling the generators as
\begin{equation}
P_\pm \longrightarrow
\sqrt{\frac{2}{\varepsilon}}
P_\pm\ ,
\qquad
Q_\pm \longrightarrow \frac{Q_\pm}{(2\varepsilon)^{1/4}}\ ,
\end{equation}
and taking the $\varepsilon\rightarrow 0$ limit. Since $\mathfrak{osp}(1|2)$ contains $\mathfrak{sl}(2)$ as a subalgebra, which correspond to the isometries of AdS$_2$, this contraction can be understood as an ordinary flat limit $\Lambda\rightarrow 0$ of Anti-de Sitter. The Maxwell superalgebra is obtained from a more subtle rescaling of the generators, given by
\begin{equation}
P_\pm \longrightarrow
\sqrt{\frac{2}{\varepsilon}}P_\pm \ ,
\qquad
Q_+ \longrightarrow \frac{Q_+}{\sqrt{\kappa}(2\varepsilon)^{1/4}}\ ,
\qquad
Q_- \longrightarrow
\sqrt{\kappa}\frac{Q_-}{(2\varepsilon)^{3/4}}
\qquad
J \longrightarrow
J+\frac{1}{\varepsilon} I\ ,
\end{equation}
where $\kappa$ is a fixed constant and in the shift in $J$ we introduced an additional generator, the central extension $I$. After this rescaling, the $\mathfrak{osp}(1|2)$ superalgebra (\ref{eq:osp12}) becomes
\begin{equation}
\begin{aligned}
{\rm Bosonic:}& \qquad
\hspace{15pt} [J,P_\pm]=\pm P_\pm\ ,
\qquad
\hspace{19pt} [P_+,P_-]=I+\varepsilon J\ , \\[4pt]
{\rm Mixed:}& \qquad
\hspace{15pt} [J,Q_\pm]=\pm\frac{1}{2}Q_\pm\ ,
\qquad  \hspace{6pt}
[P_-,Q_+]=-\frac{\kappa}{2} Q_-
\hspace{38pt}
[P_+,Q_-]=-\frac{\varepsilon}{\kappa} Q_+ \ , \\[4pt]
{\rm Fermionic:}& \qquad
\lbrace Q_+,Q_+ \rbrace=\kappa P_+\ ,
\qquad
\hspace{12pt} 
\lbrace Q_+,Q_- \rbrace=I+\varepsilon J\ ,
\hspace{25pt}
\lbrace Q_-,Q_- \rbrace =-\frac{2\varepsilon}{\kappa} P_-\ .
\end{aligned}
\end{equation}
There are three distinct cases, depending on how $\kappa$ scales with $\varepsilon\rightarrow 0$. Setting $\kappa=1$, one finds the superalgebra (\ref{eq:MaxwellSuperalgebra}), used in the main text to construct the CJ supergravity action. The second case corresponds to take $\kappa=2\varepsilon$, which (roughly speaking) gives a superalgebra in which roles of $Q_+$ and $Q_-$ are exchanged, see the discussion below (\ref{eq:MaxwellSuperalgebra}). Finally, one can take $\kappa=\sqrt{\varepsilon}$ to recover the Maxwell superalgebra found in \cite{Soroka:2004fj}. Although this last case might appear promising for defining a supergravity theory, we have explored such theory and found the boundary physics does not admit the nice asymptotic algebra (\ref{eq:AsymptoticAlgebra1}) and boundary action (\ref{eq:BdyAction2}) one gets when using the $\kappa=1$ superalgebra instead. One can check these three cases are the only ones allowed when requiring $Q_\pm$ have spin one-half $[J,Q_\pm]=\pm \frac{1}{2}Q_\pm$ together with consistency with the Jacobi identities.

A six dimensional matrix representation for the superalgebra with $\kappa=1$ is given by
\begin{equation}
\begin{aligned}
(J)_{ij}&=-\delta_{i1}\delta_{j1}
+\delta_{i2}\delta_{j2}
-\frac{1}{2}\delta_{i4}\delta_{j4}
+\frac{1}{2}\delta_{i5}\delta_{j5}\ , 
\hspace{50pt}
(I)_{ij} =\delta_{i3}\delta_{j6}  \ ,\\[4pt]
(P_+)_{ij} & = -\delta_{i2}\delta_{j6}-\delta_{i3}\delta_{j1} \ , 
\hspace{136pt}
(P_-)_{ij} =\delta_{i3}\delta_{j2}+\frac{1}{2}\delta_{i4}\delta_{j5}\ ,\\[4pt]
(Q_+)_{ij} & =
-\frac{1}{2}\delta_{i3}\delta_{j4}
+\delta_{i5}\delta_{j6}
-\frac{1}{2}\left(
\delta_{i2}\delta_{j5}
-2\delta_{i4}\delta_{j1}
\right)
  \ ,
\hspace{23pt}
(Q_-)_{ij}  =\frac{1}{2}\delta_{i3}\delta_{j5}-\delta_{i4}\delta_{j6}  \ .
\end{aligned}
\end{equation}
This explicit representation is sometimes convenient when performing some of the computations described in the main text.

\section{Matrix Model Kernel}
\label{zapp:2}

The aim of this Appendix is to derive the matrix model kernel $K(\alpha,\alpha')$ in (\ref{eq:Kernel}) for the Gaussian Hermitian matrix model $V(M)=\frac{1}{2}M^2$ in the double scaling limit indicated through (\ref{eq:double}). This is of course a very well known fact, intimately related to the bulk universality in random matrix models.

\paragraph{Finite $N$:} For any given $N$, the matrix model kernel is given in terms of a set of functions $\psi_n(\lambda)$ 
\begin{equation}\label{eq:kerfinN}
K(\lambda,\lambda') = 
\sum_{n=0}^{N-1}\psi_{n}(\lambda)\psi_n(\lambda')\ .
\end{equation}
These functions are obtained from a family of monic polynomials $P_n(\lambda)=\lambda^{n}+\mathcal{O}(\lambda^{n-1})$
\begin{equation}
\psi_n(\lambda)=\frac{1}{\sqrt{h_n}}e^{-\frac{N}{2}V(\lambda)}P_n(\lambda)\ ,
\end{equation}
where the polynomials are orthogonal with respect to the Gaussian measure defined by the matrix model
\begin{equation}\label{eq:73}
\int_{-\infty}^{+\infty}d\lambda\,e^{-\frac{N}{2}\lambda^2}P_n(\lambda)P_m(\lambda)=h_n\delta_{n,m}\ ,
\end{equation}
with $h_n$ their norm. To find an explicit expression for the polynomials, consider the generating function
\begin{equation}\label{eq:generating}
G(z)=\sum_{n=0}^{\infty}\frac{P_n(\lambda)}{n!}z^n\ .
\end{equation}
Using that the polynomials themselves satisfy the recursion relation $\lambda P_n=P_{n+1}+\frac{n}{N} P_{n-1}$ (e.g. see Appendix C of \cite{Rosso:2021orf}), the generating function can be computed as
\begin{equation}
\lambda G(z)=G'(z)+\frac{z}{N} G(z)
\qquad \Longrightarrow \qquad
G(z)=e^{z\lambda-\frac{z^2}{2N}}\ ,
\end{equation}
where the $\lambda$-dependent integration constant is fixed by requiring the polynomials are monic. Using the residue theorem we solve for each of the terms in the expansion, obtaining in this way an integral expression for the polynomials
\begin{equation}\label{eq:polyint}
P_n(\lambda)=\frac{n!}{2\pi i}\oint_{\mathcal{C}} 
\frac{dz}{z^{1+n}}e^{z\lambda-\frac{z^2}{2N}}\ ,
\end{equation}
where $\mathcal{C}$ is a contour around the origin of the $z$-complex plane. These polynomials turn out to be the Hermite polynomials, normalized such that $h_n=\frac{\sqrt{2\pi}n!}{N^{n+1/2}}$.

The polynomials in (\ref{eq:polyint}) allow us to explicitly write down the functions $\psi_n(\lambda)$ and obtain the matrix model kernel (\ref{eq:kerfinN}) for any finite value of $N$, which using the recursion relation below (\ref{eq:generating}), can be written as
\begin{equation}\label{eq:kerInt}
K(\lambda,\lambda')=
\frac{\psi_N(\lambda)\psi_{N-1}(\lambda')-\psi_{N-1}(\lambda)\psi_N(\lambda')}{\lambda-\lambda'}\ .
\end{equation}
To take the double scaling limit, it will be convenient to derive a differential equation satisfied by $\psi_n(\lambda)$. Using $P'(\lambda)=nP_{n-1}(\lambda)$ and the recursion relation one finds
\begin{equation}\label{eq:Diff}
\frac{1}{N^2} \psi_n''(\lambda)
=
\bigg[
\left(\frac{\lambda}{2}\right)^2
-
\frac{1}{N}
\left(n+\frac{1}{2}\right)
\bigg]
\psi_n(\lambda)\ ,
\end{equation}
where appropriate initial conditions can be obtained by using (\ref{eq:polyint}) to evaluate the functions and their first derivative at $\lambda=0$
\begin{equation}\label{eq:77}
\begin{aligned}
\psi_{2n}(0) & =
\frac{(-1)^n}{\sqrt{h_{2n}}}
\frac{(2n)!}{n!(2N)^n}\ ,
\qquad \qquad
\psi_{2n+1}(0)=0\ ,\\[4pt]
\psi_{2n}'(0) & =0\ ,
\hspace{110pt}
\psi_{2n+1}'(0)=\frac{(-1)^n}{\sqrt{h_{2n+1}}}\frac{(2n+1)!}{n!(2N)^n}
\ .
\end{aligned}
\end{equation}

\paragraph{Double Scaling Limit:} We now take the double scaling limit, given by
\begin{equation}
\frac{1}{N}=\hbar \delta\ ,
\qquad \qquad
\lambda_i=\alpha_i\delta\ ,
\end{equation}
with $\delta\rightarrow 0$. To compute the kernel (\ref{eq:kerInt}) in this limit, we only need the function $\psi_{N+n_0}(\lambda)$ with $n_0$ a fixed integer that does not scale with $\delta$. The differential equation (\ref{eq:Diff}) for $\psi_n(\lambda)$ evaluated at $n=N+n_0$ becomes
\begin{equation}
\hbar^2\partial_\alpha^2 \psi_{N+n_0}(\alpha)
+
\psi_{N+n_0}(\alpha)
=\mathcal{O}(\delta)\ .
\end{equation}
In the double scaling limit the terms on the right-hand side vanish and this becomes the equation for a classical harmonic oscillator. Fixing the integration constants through the double scaled version of the initial conditions in (\ref{eq:77}), one finds the solution is given by
\begin{equation}
\lim_{\delta\rightarrow 0}
\psi_{N+n_0}(\alpha)=
\frac{1}{\sqrt{\pi}}
\cos\left[\frac{\alpha}{\hbar}
+(N+n_0)\frac{\pi}{2}
\right]\ .
\end{equation}
This, together with (\ref{eq:kerInt}), allows us to obtain the final expression for the kernel in the double scaling limit
\begin{equation}
\lim_{\delta \rightarrow 0}K(\lambda,\lambda')=
\frac{1}{\pi \delta}
\frac{\sin\left[(\alpha-\alpha')/\hbar\right]}{\alpha-\alpha'}\ .
\end{equation}
Rescaling the left hand side by $\delta$, the kernel computes observables of the matrix model in the usual way, but replacing the original matrix $M$ by $\bar{M}=M/\delta$. The rescaled kernel shall be simply denoted as $K(\alpha,\alpha')\equiv\lim_{\delta\rightarrow 0} K(\lambda,\lambda')\delta$.

\addcontentsline{toc}{section}{References}
\bibliography{References}
\bibliographystyle{JHEP}

\end{document}